\documentclass[useAMS,usenatbib]{mnras}

\usepackage{graphicx}
\usepackage{color}
\usepackage{journal_names}
\usepackage{subfig}
\usepackage{float}
\usepackage{longtable}
\usepackage{multicol}  
\usepackage{rotating}
\usepackage{dcolumn}
\usepackage{lscape}  
\usepackage{amssymb}

\usepackage{pifont}
\usepackage{natbib} 

\title[OmegaWhite Survey for Cluster Variables]
  {The OmegaWhite Survey for Short Period Variable Stars VI. Open Clusters }
\author[R. Toma et al.]
  {R.~Toma$^{1,2}$\thanks{Email: ruxandra.toma85@gmail.com},
  G.~Ramsay$^2$,
  C.S.~Jeffery$^2$,
  S.A.~Macfarlane$^{3,4}$,
  P.~Woudt$^4$,
  P.J~Groot$^3$
  \\
  $^1$The Astronomical Institute of the Romanian Academy, 
      Str. Cutitul de Argint 5, Sector 4, 040557,
      Bucharest, Romania \\
  $^2$Armagh Observatory and Planetarium, 
      College Hill, 
      Armagh, 
      BT61 9DG,
      Northern Ireland, UK  \\
  $^3$Department of Astrophysics/IMAPP, 
      Radboud University, 
      P.O. Box 9010,
      6500 GL Nijmegen
      The Netherlands \\
  $^4$Department of Astronomy, 
      University of Cape Town, 
      Private Bag X3,
      Rondebosch 7701, 
      South Africa } 
\date{Accepted 2022 March 17. Received 2022 March 17; in original form 2021 November 2}

\pagerange{\pageref{Cat}\pageref{firstpage}--\pageref{lastpage}} \pubyear{2021}

\def\LaTeX{L\kern-.36em\raise.3ex\hbox{a}\kern-.15em
    T\kern-.1667em\lower.7ex\hbox{E}\kern-.125emX}

\begin{document}
\outer\def\gtae {$\buildrel {\lower3pt\hbox{$>$}} \over 
{\lower2pt\hbox{$\sim$}} $}
\outer\def\ltae {$\buildrel {\lower3pt\hbox{$<$}} \over 
{\lower2pt\hbox{$\sim$}} $}

\label{firstpage}

\maketitle

\begin{abstract}
Using light curves with $\sim$3 min cadence and a duration of 2 hrs made
using the OmegaWhite survey, we present the results of a search for short-period variable stars in the field of 20 open clusters. We identified 92 variable stars in these fields. Using a range of cluster
member catalogues and Gaia EDR3 data, we have determined that 10 are cluster members and 2 more are probable
members. Based on their position on the Gaia HRD and their photometric
periods, we find that most of these are $\delta$ Sct stars. We
obtained low-resolution optical spectroscopy of some of these cluster
members and field stars. We discuss the cluster variable stars in the
context of $\delta$ Sct stars in other open clusters.

\end{abstract}


\section{Introduction}

Open star clusters (OCs) contain a few tens to over a thousand stars which
are gravitationally bound and form and evolve in the Galactic
disc. They are important components of the Milky Way and efficient
tools for studying it.  For example, young OCs have been
used to map the Galactic rotation curve, determine the structure of
spiral arms, study mechanisms of star formation and test stellar
evolution models, whilst the structure and evolution of our Galaxy
have been traced from the distribution of old OCs
\citep{Friel-opencls1995}.  Moreover, since the cluster members are
resolved, they can be studied individually, and hence, the cluster
parameters (such as distance and age) can be determined with a higher
accuracy for OCs than for single field stars. Consequently, a study of
the population of OCs can trace the large scale structure of our own
Galaxy (e.g. \citealt{Piskunov-popGOCls-2006}, \citealt{Wu2009}).
Additionally, since the stars in an OC formed at the same time and
hence have the same age and metallicity, OCs have been used for studying
stellar evolution \citep[e.g.][]{Vanderberg1985}.

Because the age of OCs can be determined, it is interesting to verify if the variable stars properties alter as a function of age. \cite{Gilliland-M67-1991} studied
stars in the core of the old OC M67 searching for
solar-like $p$ mode oscillations. Although they did not find any such
oscillations, a set of new variable stars was found. Similarly,
\citet{HartmanVartoolsI-2008} conducted the Deep MMT (formerly Multiple Mirror Telescope) transit survey of
the intermediate-age OC M37, aiming to find transiting
exoplanets. They found 1445 variable stars, almost all being new
discoveries, many being rapidly rotating stars, but also including a sample of
pulsating stars.

Since these earlier surveys, the study of OCs has been transformed
through space missions such as {\sl Kepler}, {\sl K2}, {\sl TESS} and
wide field ground based surveys. For instance, using {\sl K2} data,
\citet{Rebull2016} determined the rotation period of stars in OCs of
different ages and hence better understand how stellar rotation period
changed over time. {\sl TESS} has also been used to search for
variability of stars in OCs and search for exoplanets around stars in
OCs \citep[e.g.][]{Bouma2019}. Although the {\sl Next-Generation Transit Survey (NGTS)} has
a main goal of discovering exoplanets, it has also been used to
determine the rotation rates of stars in open clusters
\citep[e.g.][]{Gillen2020}.

The OmegaWhite (OW) Survey was a wide field high-cadence survey whose prime aim was to
identify short period variable stars in the Galactic plane. The main
goal was to find rare ultra-compact binary systems such as the AM CVn
stars \citep{SolheimAMCVns2010} which have orbital periods of 5 --
70 min. The strategy and the pipeline of the OW survey were
outlined in \citeauthor{Macfarlane_OW-Paper1-2015} (\citeyear{Macfarlane_OW-Paper1-2015}, hereafter Paper 1),
with a detailed analysis of a large data set containing images of 134
deg$^2$ in \citeauthor{Toma-OW-Paper2-2016} (\citeyear{Toma-OW-Paper2-2016}, Paper 2). A number of
papers have followed which report populations of sources or individual
short period variables. Additionally, \citeauthor{Macfarlane-OW-Paper3-2017} (\citeyear{Macfarlane-OW-Paper3-2017}, Paper 3) showed that followup photometry of 
variable stars identified in the OW survey were confirmed with the same period, proving that the OW strategy is successful. In this
paper we report on a study which aimed to identify short period
variable stars in OCs.

\section{Overview of the OmegaWhite Survey}

Full details of the OW survey and the data reduction process are
outlined in Papers 1 and 2. Briefly, the OW survey used
OmegaCAM \citep{KuijkenOmegaCAM2011} mounted on the
ESO VLT Survey Telescope (VST, \citealt{CapaccioliVST2011}). A set of
images with exposure times of 39 s (with a mean cadence of 2.7 min)
were taken of a 1 deg$^2$ field for 2 hr. Fields were within 10 deg of
the Galactic plane. Photometry was obtained for stars in the field
using difference imaging \citep{WozniakDIAPL2000}. Variable stars were
initially identified using the Lomb Scargle periodogram
(\citealt{Lomb1976}, \citealt{Scargle1982}) and a manual verification
phase was done to vet these variable candidates. Colour information
derived from $ugriH\alpha$ filters was also obtained for many fields
using the VPHAS+--DR2 catalogue \citep{Drew-VPHAS-DR2cat-2016}. The
data for this study use the same fields analysed in Paper
2 (i.e. 134 deg$^2$, observed between 2011-2015).

\section{Variable stars in Open Cluster fields}
\label{best-cat}

Determining which stars in the field of an OC are cluster members is a
difficult and often protracted
task. \citet{Kharchenko-MWSCSurveyI-2012,Kharchenko-MWSCSurvey2013}
used the catalogue of \citet{Roeser2010} which provided 2MASS
photometry and proper motions for 900 million stars to identify more
than 3000 clusters and determine their key properties. They also give the
probability that individual stars are cluster members.  A
similar study used UCAC4 data to determine the membership of stars in
1876 OCs \citep{Sampedro2017}.  Since these studies, Gaia is
now a key source for parallax and proper motion information
\citep{GAIA-DR2-2018,Gaia2021}. This allowed
\cite{Cantat-Gaudin-OCcat-GAIA-DR2-2018} and \cite{Cantat-Gaudin-OCcat-GAIA-DR2-2020}
to determine cluster properties and to give
probabilities of whether a star is a cluster member. We now discuss how
we found variable stars in the cluster fields.

\subsection{Identifying variable stars}
\label{cls-data}

We cross-matched the Milky Way Star Clusters Catalogue (MWSC, \citealt{Kharchenko-MWSCSurvey2013}) with the OW on-sky footprint 
outlined in Paper 2. We identified 20 OCs whose angular radius (see
Table~\ref{finalsetcls}) fully overlapped the OW survey. 
These are young, intermediate and old OCs, spaning a wide range of ages between 1.0 Myr and 2 Gyr.
Although almost all of our 20 OCs have unknown metallicity, NGC 6583 has an abundance of [Fe/H] = $+0.370 \pm 0.03$ in the MWSC catalogue \citep{Kharchenko-MWSCSurvey2013} and a value of [Fe/H] = $+0.370 \pm 0.04$ reported by \cite{Heiter-Metallicity-survey2014} and  \cite{Netopil-Metallicity-survey-2016}; it is also one of the most metal rich clusters studied \citep{Magrini-NGC-6583-2010}.

We initially selected all the stars which were within the angular diameter of each cluster, and then selected those that were identified as candidate variable stars using the procedures
set out in Paper 2. Light curves were obtained for 47422 stars and 217 candidate variable stars (0.45 percent) were identified. These candidates were then
passed through our manual verification stage (Paper 2).
We are confident that 92 of these are bona fide variable stars, for which details and light curves are shown in Appendix~\ref{Cat} (Table~\ref{realvars}) and Appendix~\ref{allvars_lcs} (Figures~\ref{FIG:lcs1},~\ref{FIG:lcs2},~\ref{FIG:lcs3},~\ref{realvars-eso430-18-1},~\ref{realvars-eso430-18-2}), respectively. Examples of spurious variables include stars which had diffraction spikes rotating through their point spread function over the 2 hr of observation (see Paper 2 for further details). 
We also note that since we use the Lomb Scargle periodogram to automatically detect variable stars, the period we determine is less certain for periods which are comparable to the duration of the light curve.

\subsection{Variable stars as cluster members}
\label{vars-as-cls-members}

The next stage sets out to determine which of the variable stars we have identified in the field of OCs were {\sl bona fide} cluster members. We have used a number of catalogues from the literature and incorporate recent data derived from Gaia. 

\cite{Kharchenko-cls-par-memb-probs2005}
define a probability of a star being a cluster member as the measure
of the deviation from the mean proper motion of the cluster. They
give three types of probabilities: {\it the most probable} cluster members
are stars with a kinematic probability, $P_{kin}$ $\geq61$ percent
(i.e. stars that deviate within less than 1$\sigma$ from the mean).
Those with probabilities within $14 \leq$ $P_{kin}$ $< 61$ percent are
only {\it possible members}, in the interval (1$\sigma$,
2$\sigma$). The remaining are {\it field stars} not associated with the OC.  
  
\citeauthor{Cantat-Gaudin-OCcat-GAIA-DR2-2018} (\citeyear{Cantat-Gaudin-OCcat-GAIA-DR2-2018}, \citeyear{Cantat-Gaudin-OCcat-GAIA-DR2-2020}) determine a probability, $P_{memb}$, that a star is member of an OC based on their Gaia DR2 parallax and proper motion data. We also use the membership catalogue of \citet{Sampedro2017} which is based on UCAC4 data and have three `yes'  or `no' flags based on different criteria for membership.

Finally we used Gaia EDR3 data \citep{Gaia2021} to compare the
distance of each star with the cluster distance. We used the photogeometric distance determined by \citet{BailerJones2021} and also the 16th and 84th percentile of the
photogeometric distance as the spread of distance. For the distance
of each cluster, we assumed an uncertainty of 20 percent given in
\citet{Kharchenko-MWSCSurvey2013}; they quote an external
uncertainty on the distance of 11 percent but no internal error. Those
stars which were clearly closer or more distant than the OC (for instance many stars in the field of ESO 430-18 were significantly more distant than the cluster distance of
870 pc), we classified them as {\it not cluster members}. For the cases which were more marginal, we classified them as {\it probably not members}.

With these different criteria to hand, we assessed whether a star was a
cluster member or field star. We classed a star as a {\it cluster member} if all our criteria indicated it was a real member. For those cases where the majority of the criteria suggested cluster membership, we indicate in Table
\ref{realvars} a {\it `Probable'} membership. In some cases, there was a lack of
evidence to give a more conclusive indication of membership
probability. 
In the case of the shortest period star we found, the 29.8 min $\delta$ Sct OW
J180753.62-220904.4 (see Figure~\ref{ascc93-spec}
and Table~\ref{TAB:spectralID}), there is some uncertainty on the age of the OC
ASCC 93, (1.3 Myr given by \citealt{Kharchenko-MWSCSurvey2013}, 16 Myr by
\citealt{Sampedro2017}). However, since $\delta$ Sct stars are older than the age of this cluster, 
this star is not associated with the OC ASCC 93.

Out of the 92 variable stars, we identified 10 as cluster members and other 2 as probable members. We classified 6 as probably not members, 68 as not cluster members and the remaining 6 stars have unknown membership status. To summarise, 12 out of 92 variable stars have the highest probabilities to be physically members of open clusters, and we consider them as {\it cluster members} during the rest of this paper. 

\subsection{Variable stars on the Gaia HRD}

\begin{figure}
\centering
\includegraphics[scale=0.65]{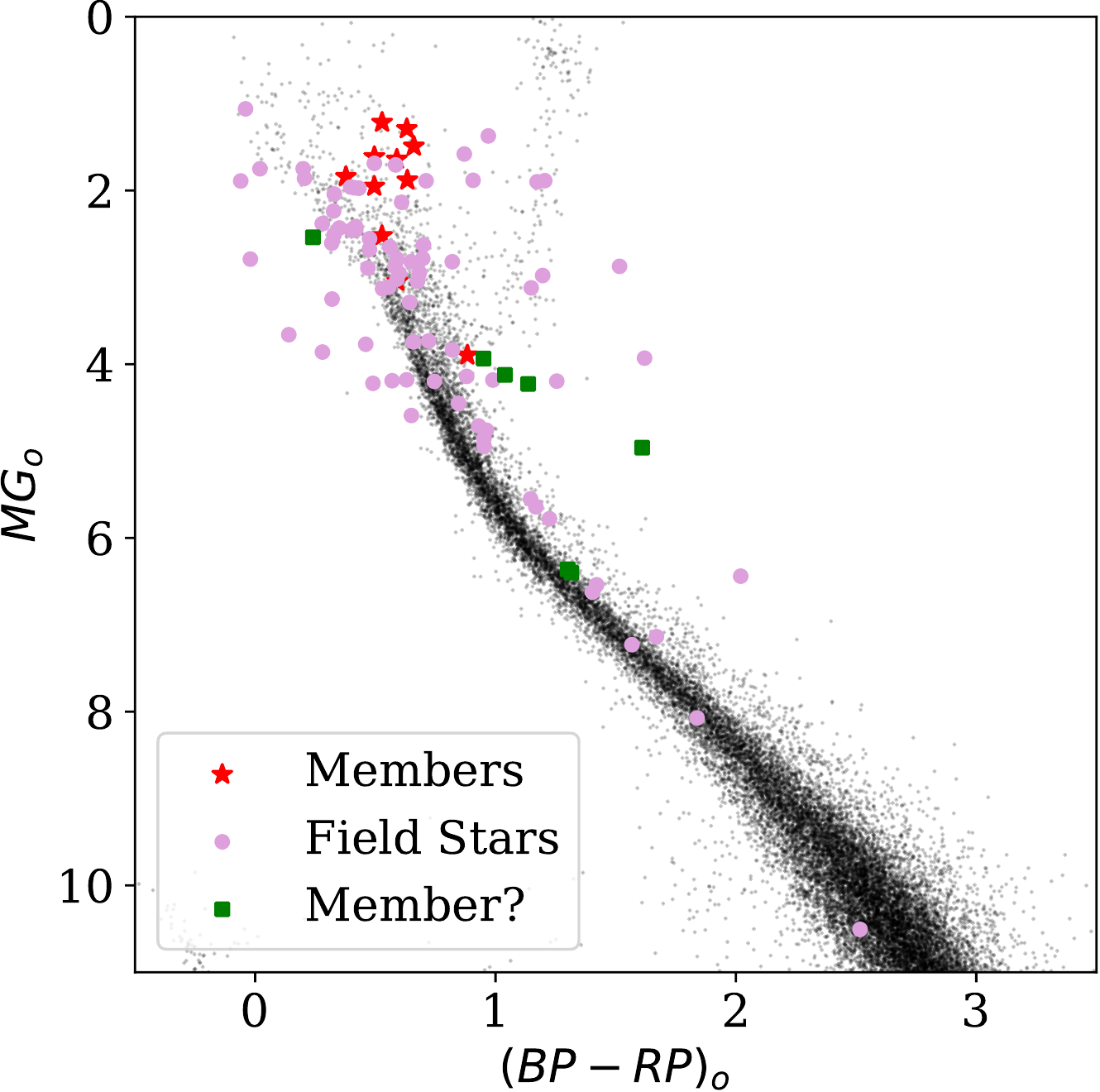}
\caption{The position of variable stars in the region of open
  clusters in this study in the Gaia $(BP-RP)_{o}$, $MG_{o}$ plane. We
  differentiate those which we regard as cluster members and probably
  members (shown as red stars), those we regard as field stars or probable field stars (pink dots)
  and those for which we are unable to determine whether they are members
  or not (green squares). The smaller dark dots are stars within 50 pc of the Sun which we
  assume are not affected by reddening.}
\label{gaiahrd}
\end{figure}

We also used Gaia EDR3 to place the variable stars on the Hertzsprung-Russell Diagram (HRD), again
using the photogeometric most likely distance determined by
\citet{BailerJones2021}. We then deredden the $(BP-RP)$, $M_{G}$
values using the 3D-dust maps derived from Pan-STARRS1 data
\citep{Green2019} and the relationship between $E(B-V)$ and $E(BP-RP)$
and $A_{G}$ and $E(BP-RP)$ outlined in \citet{Andrae2018}. For those
stars just off the edge of the Pan-STARRS1 field of view (stars with
$\delta<-30^{\circ}$) we take the nearest reddening distance
relationship. We show the resulting Gaia HRD along with stars within
50 pc, which we assume have no significant reddening, in Figure
\ref{gaiahrd}. We find that the majority of stars which are variable
and cluster members have likely evolved off the main sequence implying
that the variability is due to stellar pulsations. Field variable
stars are located close to the main sequence (down to early M type),
evolved off the main sequence, but also blueward of the upper main
sequence which may indicate they contain evolved stars. (Due to fields
being close to the Galactic plane and due to the nature of how we
deredden the $(BP-RP)$, $M_{G}$ values, there is some uncertainty in
their location on the Gaia HRD).

\begin{table*}
\centering
\caption[]{{\footnotesize The 20 open clusters in this study. The
    first two columns give the ID number in the MWSC catalogue
    \citep{Kharchenko-MWSCSurvey2013} and the cluster name. They are
    followed by the ESO semester when the data was taken in and the
    internal OW field name. Next, the sky coordinates (J2000) of the
    cluster centre, the angular radius of the cluster extent, the
    distance, the $E_{B-V}$ and the cluster age are all taken from the MWSC. 
    Last, the number of variable stars found in the angular area of each cluster is given.
    An asterisk indicates the clusters that were found in \citeauthor{Cantat-Gaudin-OCcat-GAIA-DR2-2018} (\citeyear{Cantat-Gaudin-OCcat-GAIA-DR2-2018}, \citeyear{Cantat-Gaudin-OCcat-GAIA-DR2-2020})'s catalogues. }}
\label{finalsetcls}
\begin{tabular}[pos]{llcrllrrlll}
\hline
  \multicolumn{1}{c}{MWSC} &
  \multicolumn{1}{l}{Cluster} &
  \multicolumn{1}{c}{ESO} &
  \multicolumn{1}{c}{OW} &
  \multicolumn{1}{c}{RA} &
  \multicolumn{1}{c}{Dec} &
  \multicolumn{1}{c}{r} &
  \multicolumn{1}{c}{d} &
  \multicolumn{1}{c}{E$_{B-V}$} &
  \multicolumn{1}{c}{log t } &
  \multicolumn{1}{c}{\# Variable stars}\\
ID  & Name & Sem & Field & (deg) & (deg) & (arcmin) & (pc) & (mag) & (yr) & \\
 \hline
1392 &   Turner 12      &  88   &    3a     &      119.5 & --29.3  & 7.5 & 4500 & 0.73 & 8.5 & 0 \\ 
1420 &   ESO 430-09           &  88   &    3b     &      120.6    & --29.8  & 4.8  & 2270 & 0.21 & 8.5 & 2  \\ 
1432 &   NGC 2533$^{*}$      &  88   &    4a     &      121.8 & --29.9   & 8.4  & 2960 & 0.46 & 8.4 & 6 \\ 
1345 &   Teutsch 25           &  88   &    11a    &      117.1 & --27.9    & 4.2  & 2840 & 0.42 & 9.3 & 0 \\ 
1383 &   NGC 2483             &  88   &    12a    &  118.9 & --27.9 & 6.9 & 1650 & 0.35 & 7.6 & 7    \\
2668 &   ESO 520-20           &  91   &    17b    &      265.9 & --24.7  & 4.5 & 2040 & 2.33 & 9.2 & 2 \\ 
2690 &   Dutra--Bica 44 &  91   &    18a    &    266.6    & --24.9 & 4.2 & 2260 & 2.41 & 9.1 & 0 \\
2791 &   Teutsch 14a$^{*}$   &  93   &    85a    &      270.9 & --22.1  & 9.6  & 2110 & 1.50 & 8.5 & 7 \\ 
2814 &   ASCC 93              &  93   &    85b    &      272.0   & --22.3   & 9.9 & 1830 & 0.58 & 6.1 & 1  \\ 
2845 &   NGC 6573             &  93   &    86a    &      273.4 & --22.1   & 5.7 & 3030 & 1.06 & 8.8 & 3  \\ 
2860 &   NGC 6583$^{*}$      &  93   &    86b    &      274.0  & --22.1  & 6.0   & 1750 & 0.56 & 9.0 & 9  \\ 
2708 &   Dutra--Bica 51 &  93   &   100b    &   267.4 & --31.3 & 3.3 & 1350 & 3.12 & 6.0 & 0 \\
2613 &   Antalova 3           &  93   &   103a    &      262.7 & --32.2  & 6.6  & 2590 & 1.35 & 8.4 & 1 \\ 
1273 &   Haffner 11     &  94   &    34b    &   113.8 & --27.7 &  6.0 & 5160 & 0.42 & 8.9 & 0 \\
1426 &   ESO 430-14           &  94   &    40b    &      120.9 & --31.4  & 7.8  & 2860 & 0.83 & 7.2 & 15 \\ 
1391 &   FSR 1342$^{*}$     &  94   &    40a    &      119.5 & --31.5  & 7.2  & 3630 & 0.94 & 8.3  & 2 \\ 
1208 &   Ivanov 6       &  94   &    107a   &    111.1 & --24.6  & 6.0 & 2430 & 1.15  &  6.6 & 0 \\ 
1279 &   Riddle 5             &  94   &    108b   &      114.2 & --24.7  & 3.6  & 2710 & 0.90 & 8.8 & 2 \\                     
1431 &   ESO 430-18           &  94   &    114b   &      121.8 & --30.8  & 17.4  & 870  & 0.0   & 8.9 & 32 \\    
1425 &   Ruprecht 51          &  94   &    114a   &      120.9  & --30.7   & 5.7 & 2530 & 0.56 & 8.8  & 3 \\ 
\hline
\end{tabular}
\end{table*}

\section{Spectroscopic identification of variable stars}
\label{spectraid}

As part of the OW follow-up programme, we acquired low-resolution
spectra of 12 variable stars found within the angular radius of the
OCs to help determine their nature (only three were subsequently found
to be OC members). The stars were observed between April to May 2016
using the Spectrograph Upgrade-Newly Improved Cassegrain (SpUpNIC,
\citealt{Crause2019_SpUpNIC}) mounted on the 1.9m telescope at the
South African Astronomical Observatory in Sutherland. The G7 grating
was used which has a wavelength range of 5600 \AA, 300 lines per mm
and a resolving power of $\sim$700. The observing log is given in
Table~\ref{TAB:spectra-log}.  The spectra were then reduced and flux and wavelength
calibrated using the same standard techniques described in
\citealt{Macfarlane-OW-Paper3-2017} (Paper 3). 

\begin{table*}
\centering
\caption[]{{\footnotesize The observing log of follow-up spectra
    obtained for 12 variable stars found in the angular area of 20
    open clusters that overlap OmegaWhite fields.  The survey ID,
    cluster name, date of observation, exposure time, airmass and the
    slitwidth are listed for each star. }}
\label{TAB:spectra-log}
\begin{tabular}[pos]{llllrcc}
\hline
  \multicolumn{1}{l}{OW Star ID} &
   \multicolumn{1}{l}{Cluster} &
   \multicolumn{1}{l}{Date-obs} &
  \multicolumn{1}{r}{Exp} &
  \multicolumn{1}{l}{Airmass}  &
  \multicolumn{1}{l}{Slitwidth}  \\
 & & & sec & & $''$ \\
\hline
OW J080659.21--300022.2 & NGC 2533    &  30-04-2016 & 1800 & 2.31 & 1.80 \\  
OW J080706.92--295626.7 & NGC 2533    &  30-04-2016 & 1500 & 1.83 & 1.80 \\ 
OW J080711.98--294621.7 & NGC 2533    &  02-05-2016 & 2100 & 1.19 & 1.05 \\ 
OW J080721.21--294516.6 & NGC 2533    &  01-05-2016 & 1800 & 1.08 & 1.50 \\
OW J180319.05--215901.8 & Teutsch 14a &  03-05-2016 &  900 & 1.06 & 1.50 \\
OW J180753.62--220904.4 & ASCC 93     &  01-05-2016 & 1200 & 1.04 & 1.80 \\
OW J181329.36--220502.4 & NGC 6573    &  02-05-2016 &  900 & 1.01 & 1.80 \\
OW J181531.53--220755.3 & NGC 6583    &  02-05-2016 &  900 & 1.03 & 1.80 \\
OW J080554.11--305750.0 & ESO 430-18  &  03-05-2016 & 2400 & 1.15 & 1.50 \\
OW J080603.57--305142.5 & ESO 430-18  &  30-04-2016 & 1800 & 1.49 & 2.10 \\
OW J080603.79--305050.5 & ESO 430-18  &  02-05-2016 & 1800 & 1.08 & 1.50 \\
OW J080635.21--305741.5 & ESO 430-18  &  01-05-2016 & 1200 & 1.24 & 1.50 \\
\hline
\end{tabular}
\end{table*}

\subsection{Identification of stellar types}

Based on the shape of their light curves, their periods and location
on the Gaia HRD, we expect that most of the 12 stars, which we obtained
spectra for, are likely $\delta$ Sct stars which have an A -- F spectral
type. We made a preliminary determination of the stars spectral type
by comparison with the JHC spectral atlas \citep{JHC-atlas1984}.

\begin{figure}
\centering
\includegraphics[scale = 0.17]{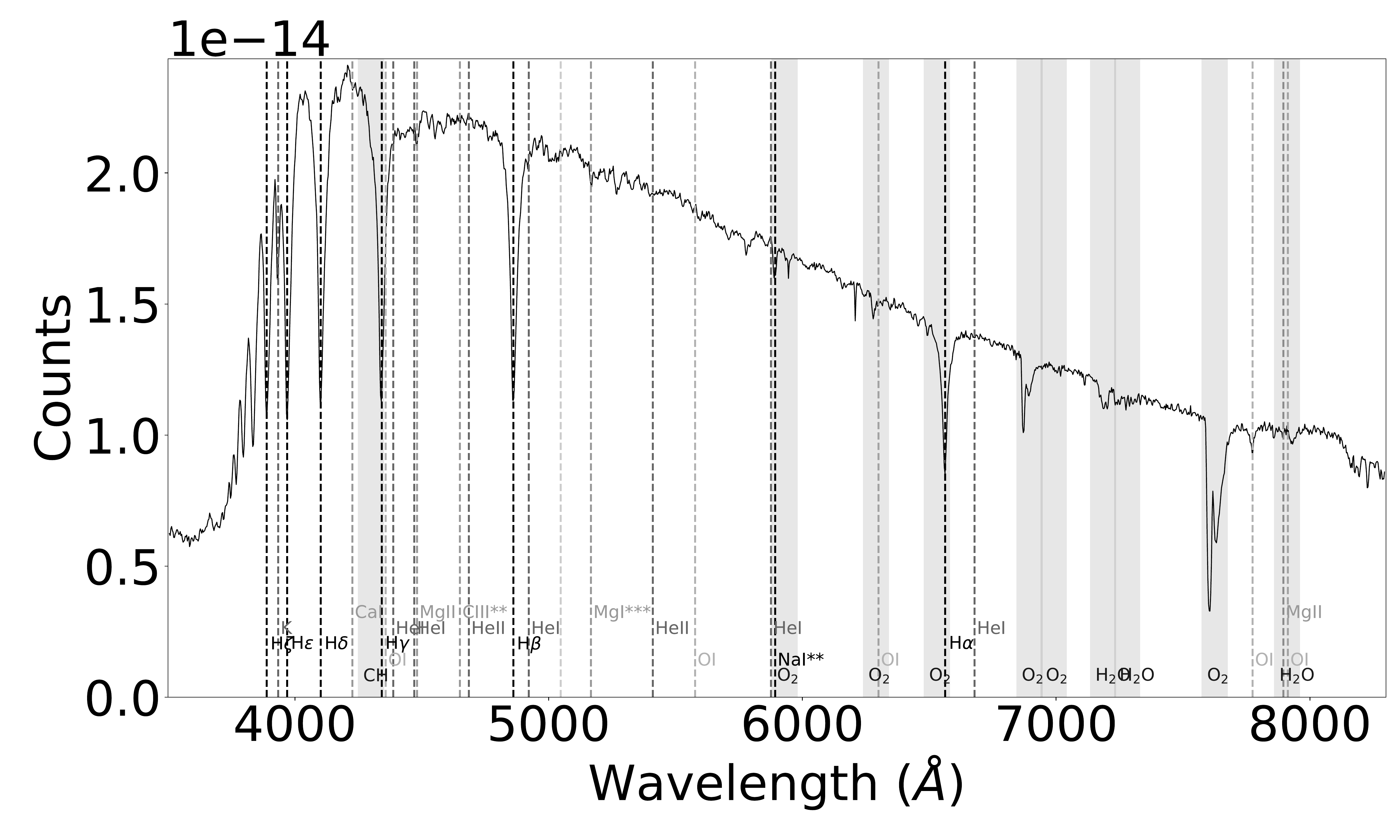}
\caption{The spectrum of the short-period $\delta$ Sct-like variable star, OW J180753.62--220904.4, which has a 66 percent chance to be a member of the
 open cluster ASCC 93. The spectrum was flux and wavelength calibrated. The dashed lines indicate spectral lines and the shaded areas atmospheric telluric bands (Papers 3, 4). The spectrum indicates that our star is an A4V type, confirming it to be a $\delta$ Sct pulsator.}
\label{ascc93-spec}
\end{figure}

The spectra of the variable stars are shown in Appendix~\ref{ap-spectra-JHC} (Fig.~\ref{spectra1}, ~\ref{spectra2}). 
In Figure~\ref{FIG:sp-avsfstars}, we compare one of our stars with A and F stars from the JHC atlas, as an example. Thus, we find that ten stars have an A spectral type because they exhibit strong H Balmer lines and weak Ca
II lines (Figure~\ref{FIG:sp-astars}); we show one example spectrum in
Figure~\ref{ascc93-spec}. Two stars, OW J080659.2-300022.2 and OW
J180319.05-215901.8, have unusual spectra, but the presence of
Hydrogen Balmer lines, indicate that these are typical A -- F stars,
possibly affected by interstellar reddening
(Figure~\ref{spectra2}).

\subsection{Identification of stellar types using the equivalent width ratios}
\label{eqwidth}

\begin{table*}
\centering
\caption[]{{\footnotesize A sample of 12 variable stars found
    overlapping the area of 20 open clusters with follow-up
    low-resolution spectra available. The survey ID, the cluster name,
    and the MWSC kinematic membership probability are given first
    \citep{Kharchenko-MWSCSurvey2013}. Next, the membership
    probability obtained from Gaia-DR2 data published by
    \cite{Cantat-Gaudin-OCcat-GAIA-DR2-2018} is listed. The empty
    spaces indicate that there were no data in the catalogue for those
    particular stars. The equivalent width ratios of the Ca II lines
    (i.e. Ca K/(Ca H+H$\epsilon$)), the estimated spectral types and the cluster membership
    are given in the last three columns. See text for details. }}
\label{TAB:spectralID}
\begin{tabular}[pos]{llrllll}
\hline
  \multicolumn{1}{l}{Star ID} &
   \multicolumn{1}{l}{Cluster} &
   \multicolumn{1}{l}{P$_{kin}$} &
   \multicolumn{1}{l}{P$_{Memb}$} &
   \multicolumn{1}{l}{EW Ratio} &
  \multicolumn{1}{l}{Spectral type } &
  \multicolumn{1}{l}{Cluster member? }\\
 &    &  (\%)  &  (\%) &  (Ca K/ Ca H+H$\epsilon$) & JHC + EW &  \\
\hline
OW J080659.21--300022.2 & NGC 2533   &  0.0  & 70.0  & 0.256 $\pm$ 0.093 & reddened A8V &  Cluster member \\
OW J080706.92--295626.7 & NGC 2533   & 78.4 & 80.0  & 0.133 $\pm$ 0.034 & A3V & Cluster member \\
OW J080711.98--294621.7 & NGC 2533   &  9.2 & 50.0  & 0.208 $\pm$ 0.029 & A7V & Field star \\
OW J080721.21--294516.6 & NGC 2533   & 66.7 & 50.0  & 0.270 $\pm$ 0.043 & A8V & Field star \\
OW J180319.05--215901.8 & Teutsch 14a & 54.4 & 10.0 & Ca lines invisible & reddened A/F & Field star \\
OW J180753.62--220904.4 & ASCC 93    & 65.5 &  & 0.141 $\pm$ 0.016 & A4V & Field star \\
OW J181329.36--220502.4 & NGC 6573   & 63.2 &  & 0.169 $\pm$ 0.012 & A5V & Field star \\
OW J181531.53--220755.3 & NGC 6583   & 14.1 & 90.0  & 0.231 $\pm$ 0.026 & A7V/A8V & Cluster member \\
OW J080554.11--305750.0 & ESO 430-18 &  0.0  &  & 0.178 $\pm$ 0.022 & A6V & Field star \\
OW J080603.57--305142.5 & ESO 430-18 & 13.1 &  & 0.143 $\pm$ 0.012 & A4V & Field star \\
OW J080603.79--305050.5 & ESO 430-18 &       &  & 0.266 $\pm$ 0.069 & A8V & Field star \\
OW J080635.21--305741.5 & ESO 430-18 &  0.0  &  & 0.073 $\pm$ 0.004 & A0V/A1V & Field star \\
\hline
\end{tabular}
\end{table*}

\begin{figure}
\centering
\includegraphics[scale=0.32]{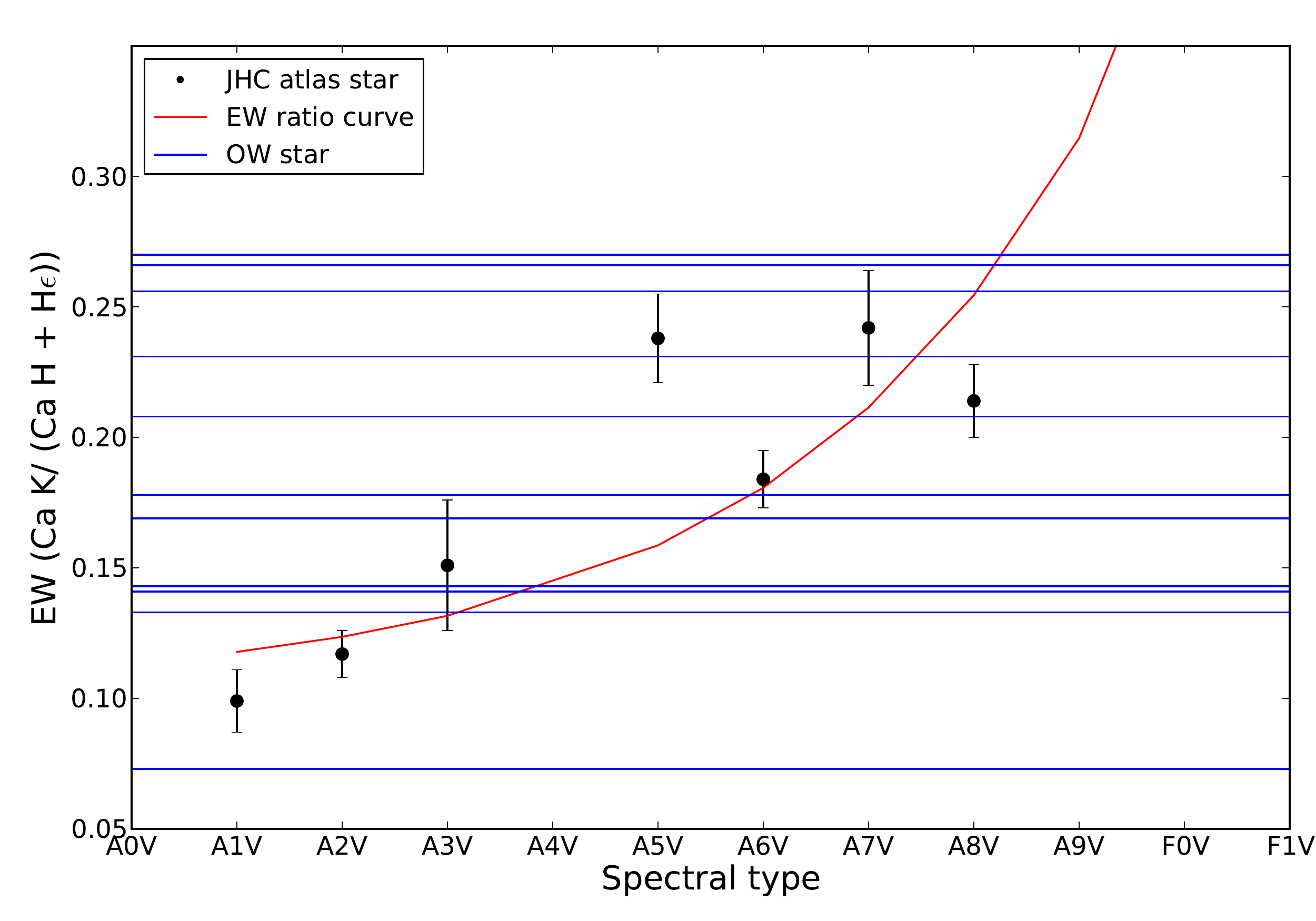}
\caption{The data points represent the measured EW ratio (Ca K/ (Ca H+
  H$\epsilon$)) for a set of reference A stars from the JHC atlas
  \citep{JHC-atlas1984}.  The EW ratio curve is obtained by fitting
  the data points with an exponential function. The measured EW ratios
  of the 11 variable stars in open clusters are displayed as
  horizontal lines.}
\label{EWplot}
\end{figure}

To refine the spectral type of the 12 variable stars we determined the
equivalent width (EW) of the Ca II line (K, 3933 \AA) and H$\epsilon$
(3970 \AA) for identification (c.f. \citealt{Jaschek1990}). We
modified the EW method used in Paper 3. For our 12 stars, we measured
the EWs of Ca K and the Ca H + H$\epsilon$ doublet because the K line
changes significantly from one A subtype to the other. We used the
interactive tool {\tt splot} in {\tt IRAF} \citep{Tody-iraf-1993} for
the measurements. We obtained a calibration curve using as reference
all the A1V up to F4V main sequence stars given by
\cite{JHC-atlas1984}.  The same two Ca lines were measured and the EW
ratios determined for these reference stars. Lastly, we fit the
measured ratios as a function of spectral type using an exponential
function. The results of these steps are shown in Figure~\ref{EWplot}. The intersection of
each line with the curve indicates the likely spectral type of our
variable stars, from A0V to A8V; see also Table
\ref{TAB:spectralID}.

In most cases, there is a good agreement between
the spectral type assigned using the JHC atlas and the EW
approach. However, one star OW J080706.92-295626.7, is identified as
an A3V star by the EW ratio, although the visual comparison places it
as an earlier A0/A1V star. Further, OW 080659.2-300022.2 is thought
to have A8V spectral type from the EW ratio, but approximately F4V
from its location in the Gaia HRD (Figure \ref{gaiahrd}).
The OW J180319.05-215901.8 (bottom spectrum in Figure~\ref{spectra2}) is so affected by reddening that the Ca lines are not visible and hence we could not measure the EW ratio for this star. Thus we cannot tell if it is an A or F star.

\section{Nature of the Variable stars}
\label{natureofvars}

\begin{figure*}
\centering
\includegraphics[scale=0.34]{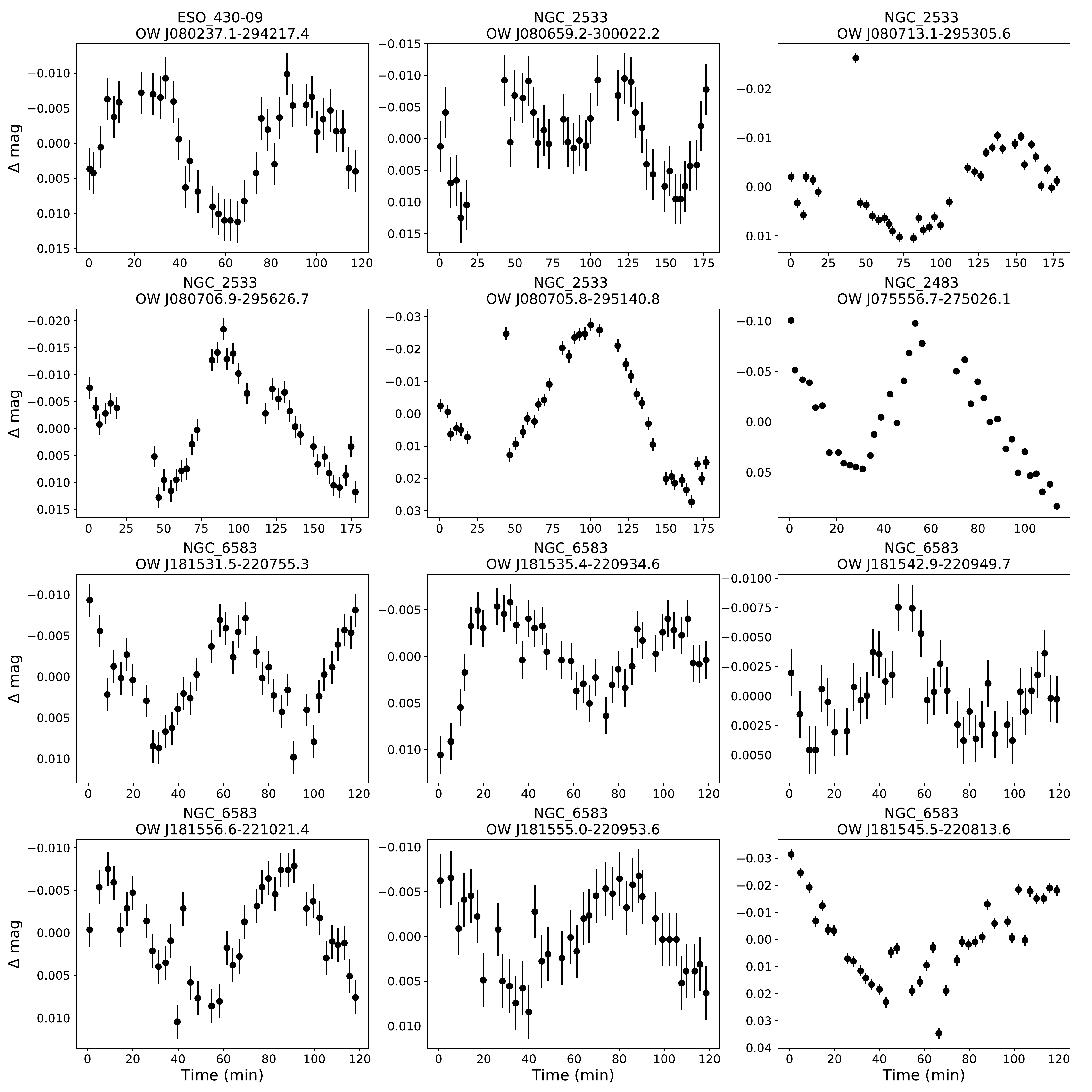}
\caption{The light curves of the 12 variable stars that we found to be members or highly probable
  members of the 20 open clusters found overlapping the set of OW fields. For each, we list the cluster name and its survey name. Further details are given in Table~\ref{clusterdeltaSct}.  }
\label{members-plot}
\end{figure*}

We show details of the 92 variable stars in Table \ref{realvars} and
their location on the Gaia HRD in Figure \ref{gaiahrd}. We have
identified 12 stars as either being cluster members or probable members - i.e.
with a high membership probability and we consider them as {\it cluster members} from now on
(their details are outlined in Table
\ref{clusterdeltaSct} and the light curves are shown in Figure~\ref{members-plot}); six are members of the cluster NGC 6583; four are members of
NGC 2533; one is a probable member of ESO 430-18 and the remaining a probable member of
NGC 2483.
Considering the shape of their light curves and colours, these cluster
stars are $\delta$ Sct pulsators (\citealt{Breger-delsct-1979}, \citeyear{BregerDelScuti2000}), although we cannot
rule out some of them being SX Phe stars, which are their metal poor counterparts \citep{NemecMateoSXPhe1990}. High-resolution spectra are necessary to study the metal lines in order to differentiate between these two types.

We obtained spectra for three stars which we consider cluster members. OW
J181531.53--220755.3 in NGC 6583 is an A7/A8V star
(Figure~\ref{FIG:sp-astars}) consistent with it being a low amplitude
$\delta$ Sct.  OW J080706.92--295626.79 in NGC 2533 is an A3V star
and the light curve looks like that of a low amplitude $\delta$ Sct
pulsator. The analysis outlined in \S \ref{eqwidth} suggested that OW
J080659.21--300022.2 (i.e. top spectrum in the Figure~\ref{spectra2})
is a A8V star, although the shape of its continuum pointed to it being
reddend. However, its position on the Gaia HRD is consistent with the spectral type F4V: both spectral types point to a $\delta$
Sct classification. Overall, the position of these variables in the
Gaia HRD indicates that they are either main sequence stars, or have
evolved off the main sequence, which is consistent with them being
$\delta$ Sct stars.

We have classified two stars as probable cluster members. 
OW J075556.7-275026.1 in NGC 2483 is a $\delta$ Sct (there has been a debate about whether NGC 2483
is a genuine cluster, e.g. \citealt{FitzgeraldMoffat1975}). It has a
position on the dereddened Gaia HRD consistent with a late F spectral
type.  OW J080237.1-294217.4 in ESO 430-09 does not have a Gaia
$(B-R)$ colour but has a dereddend MG$_{o}$ consistent with a late K
spectral type. 

Of the remaining stars, 6 have
insufficient information to assess their membership, with the
remaining 74 being not or probably not cluster members. Figure
\ref{gaiahrd} shows that many of the field variable stars (and those
with no membership information) are pulsating stars with a
smaller number being hotter than the main sequence implying they would be
evolved stars (this part of the Gaia HRD includes hot subdwarf stars, sdO/Bs, located blue-ward of the horizontal branch, around $M_{G} \sim 5$ mag, as shown by \citealt{GaiaColab2018-HRDs}, \citealt{Geier2019-hotsdII}, \citealt{DalTio2021-GaiaHRD}).

\begin{figure}
\centering
\includegraphics[scale=0.27]{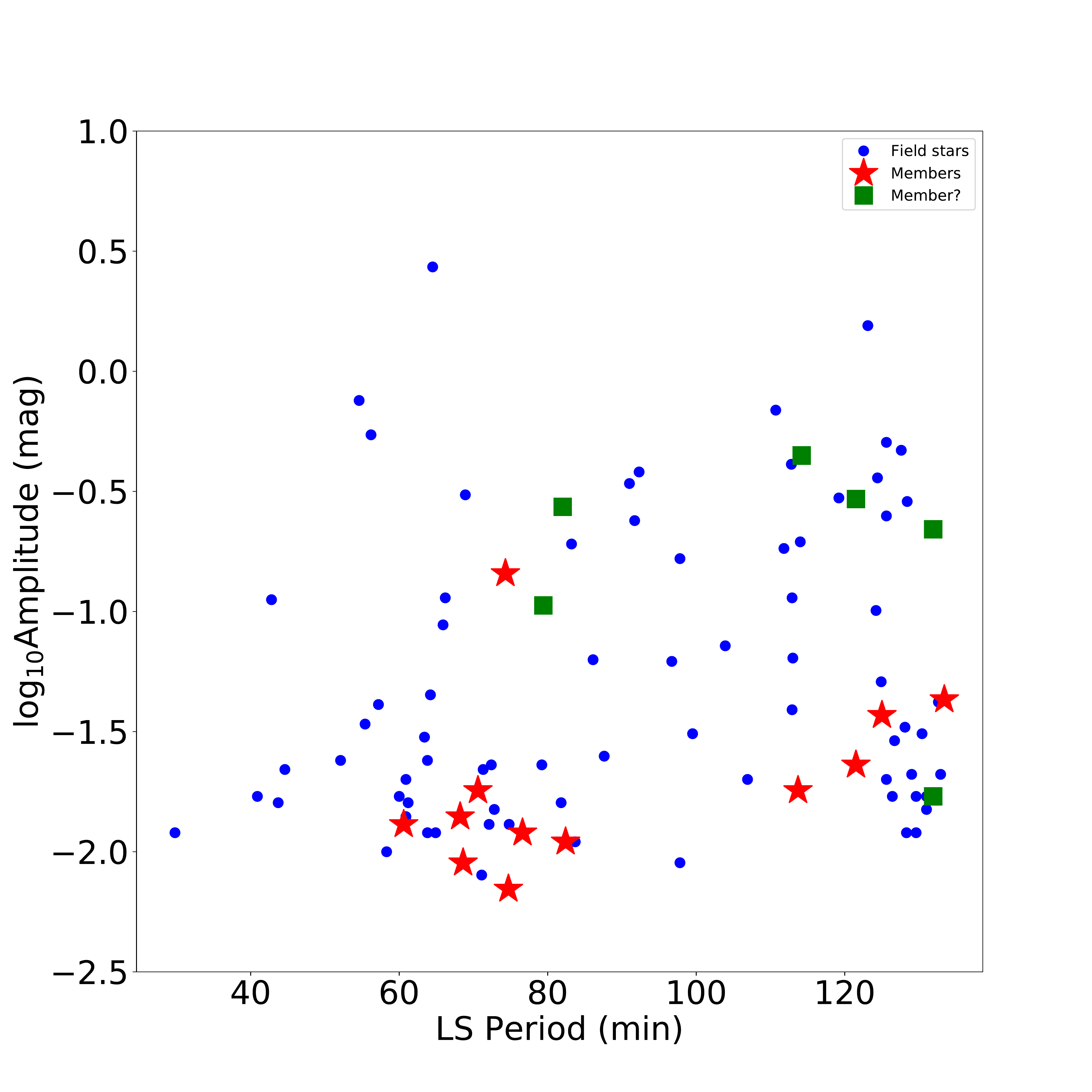}
\caption{The period-amplitude diagram of the variable stars reported here in log scale. The cluster members and probably members are shown as red stars, the field stars and probable field stars are shown as blue dots and the green squares are the stars with unknown membership status.}
\label{per-amp-plot}
\end{figure}

In Papers 2 and 3 we showed that the amplitude of the periodic modulation (which is derived using the Lomb Scargle Periodogram) could be used to help determine the nature of the variable sources. In Fig. \ref{per-amp-plot} we show the amplitude as a function of period. Of the 12 OC members, 11 have an amplitude consistent with typical low amplitude $\delta$ Sct stars. One star, OW J075556.7-275026.1, a member of NGC 2483, has an amplitude of 0.14 mag and could be a high-amplitude $\delta$ Sct star (HADS, \citealt{Rodriguez1996-HADS}, \citealt{Alcock2000-HADS}, \citealt{Garg2010-HADS}). 
We note 30 more stars showing high amplitudes that could be HADS. Of these, five stars which we cannot determine if they are cluster members might be HADS based on their amplitude and location on the Gaia HRD. The remaining are field stars.
Nevertheless, those stars with high amplitudes could also be eclipsing binary stars or compact pulsators, although we note that some stars showing high amplitudes (e.g. OW J080610.3-304739.1, Fig. \ref{realvars-eso430-18-1}) are artificially high due to some spurious data points.

\begin{figure*}
\centering
\subfloat{\label{FIG:umg-vs-gmr}\includegraphics[scale = 0.35]{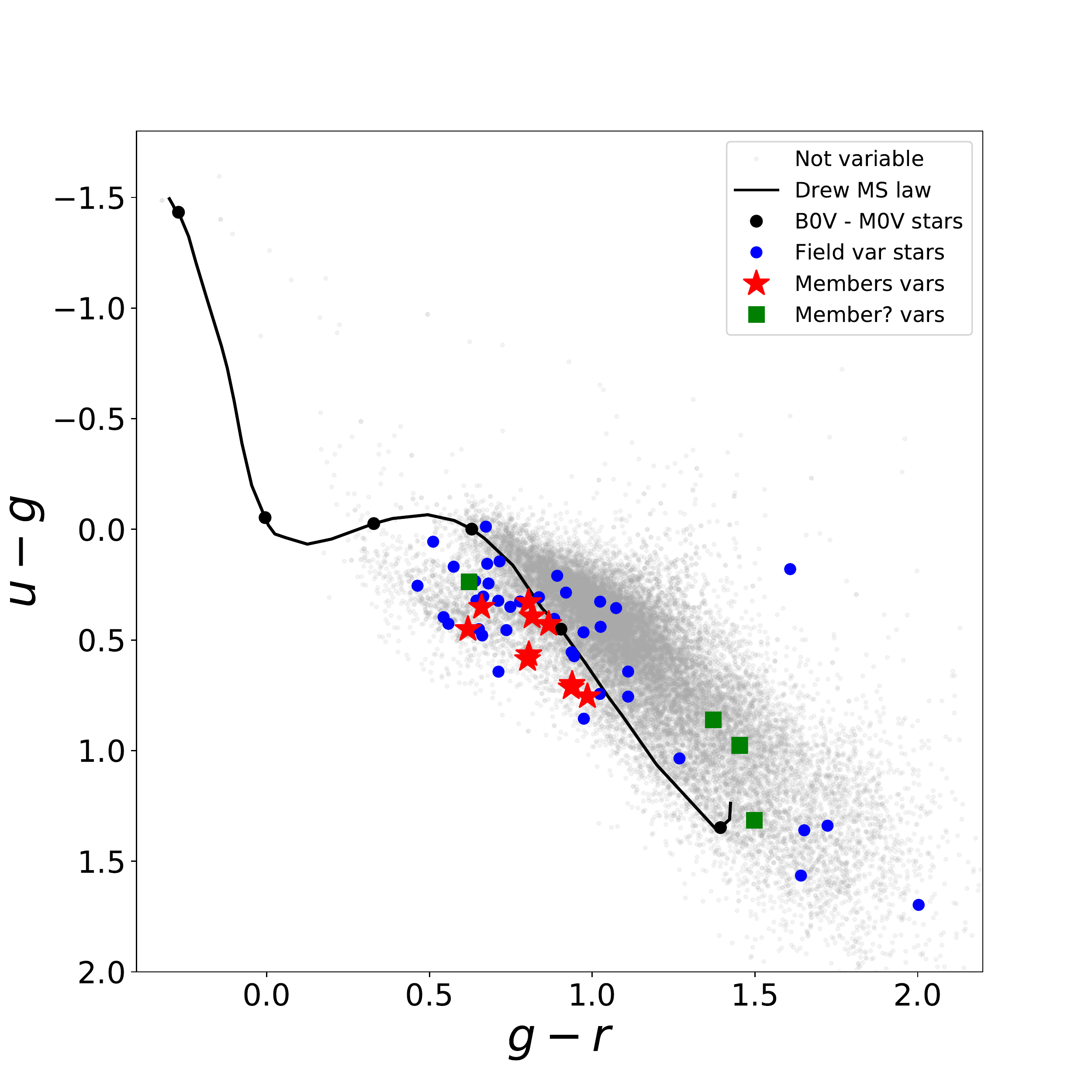}}
\subfloat{\label{FIG:rmHa-vs-rmi}\includegraphics[scale = 0.35]{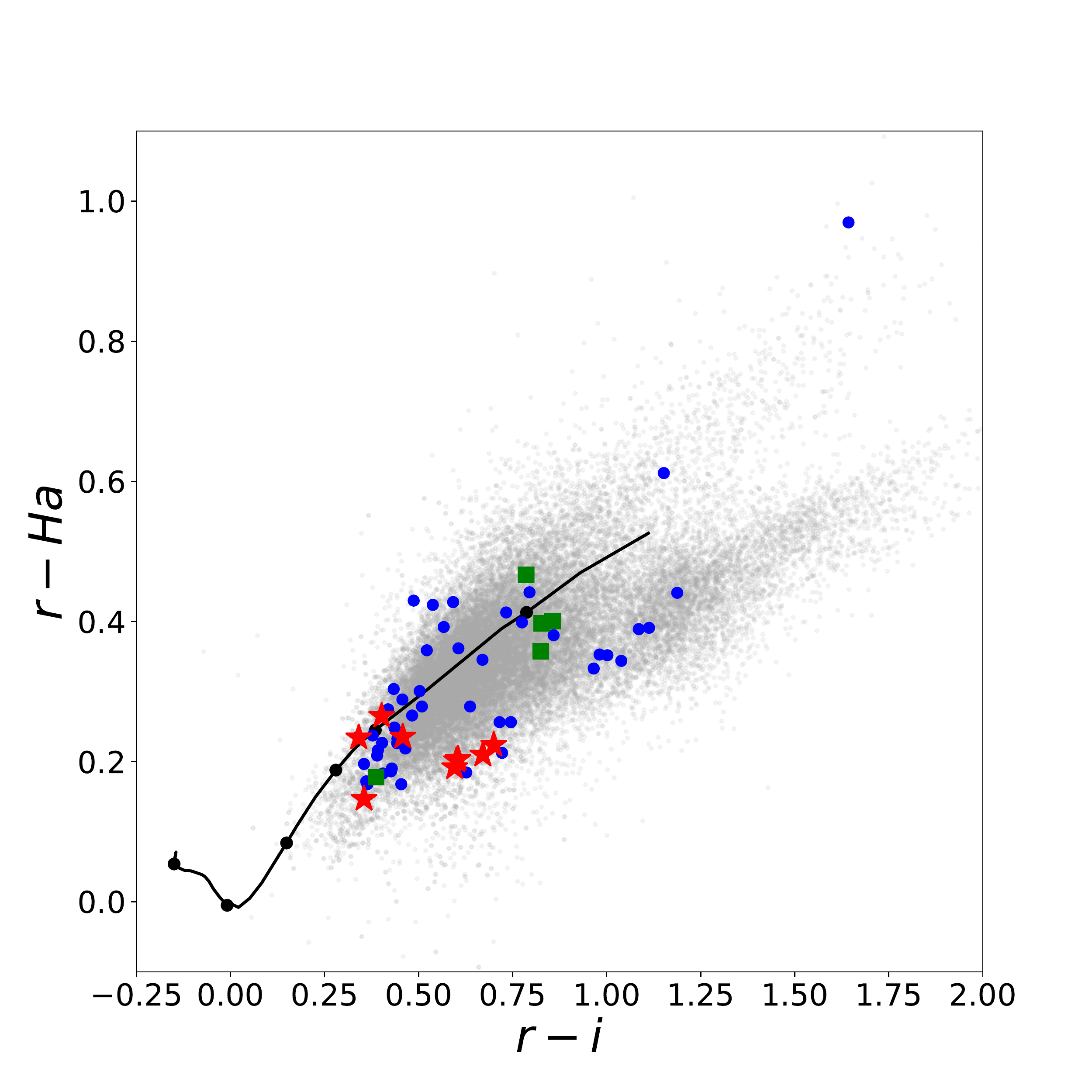}}
\caption[]{{\footnotesize The colour-colour diagrams in the $u-g$, $g-r$ (left side) and $r-H\alpha$, $r-i$ (right side) planes for the sample of variable stars reported here.  The colour information was taken from the VPHAS+-DR2 catalogue \citep{Drew-VPHAS-DR2cat-2016}. The light grey small dots are all stars measured in the areas of the clusters. The black line represents the unreddened main-sequence derived by \cite{DrewVPHAS+2014}. On this line we mark the position of B0V - M0V stars for reference. We indicate the variable stars found as cluster members as red star symbols; the green squares are stars with unknown membership probability and the blue dots are the field variable stars.}}
\label{col-col-diagrams}
\end{figure*}

We now use the VPHAS+ colour information \citep{DrewVPHAS+2014,Drew-VPHAS-DR2cat-2016} to confirm the variable star classification of the cluster member stars and identify other stars which may have interesting properties. We show the stars in the $(g-r),(u-g)$ and $(r-i)$,$(r-H\alpha$) planes in Fig. \ref{col-col-diagrams}. The cluster members have colours consistent with being reddened stars of A-F spectral type, confirming that they are $\delta$ Scutis. The redder stars at the bottom right in the $ugr$ diagram could also be eclipsing and/or contact binaries. 
One field star, OWJ 073646.1-244016.9, is significantly bluer than the main sequence and has $r-H\alpha$ = 1.0 indicating H$\alpha$ emission and therefore it is worthy of follow-up spectroscopy (it is located near the lower part of the main sequence in Fig. \ref{gaiahrd}).

\section{$\delta$ Scuti stars in Open Clusters}

\citet{Rodriguez2002}'s compilation of $\delta$
Sct stars in 22 OCs shows 84 stars with periods of typically 1--3 hrs. The age
of the OCs are widely varied, with \citet{Rodriguez2002} noting 3
$\delta$ Sct stars in the $\alpha$ Per cluster (age $\log t/{\rm y}\sim7.7$,
\citealt{Kharchenko-MWSCSurvey2013}), and 4 $\delta$ Sct stars in Melotte 71 ($\log t/{\rm y}\sim9.0$,
\citealt{Kharchenko-MWSCSurvey2013}). \citet{Chang2013} in their
statistical survey of Galactic $\delta$ Sct stars identify nearly 100 such
variables as OC members.  Using {\sl K2}, \cite{Rebull2016} report
eight pulsator-like stars in the Pleiades ($\log t/{\rm y}\sim8.2$) of
which five are clear $\delta$ Sct stars with periods $\lesssim 2.4$ hrs with
three more with similar periods $\sim7.2$ hrs.  More recently,
\cite{Durgapal-del-Sct-OCs-2020} searched for variable stars in four
OCs using a 1.0m ground based telescope and found 5 $\delta$ Sct stars, 4
in King 7 ($\log t/{\rm y}\sim8.9$) and 1 in King 5 ($\log t/{\rm y}\sim9.1$);
their periods are in the range of 3.1 - 5.0 hrs.

To compare, we found a small number of $\delta$ Sct stars, only 12, in four open clusters (NGC 2533, NGC 6538, NGC 2483 and ESO 430-09) from our sample of 20 (\S ~\ref{natureofvars}).

NGC 2533 is an intermediate age
cluster of 250 Myr old and NGC 6583 is old (1 Gyr). Although we noted
earlier the debate on whether NGC 2483 was a real OC, its age appears
to be young (40 Myr). The periods we found for these cluster members
are between 60 - 125 min.

The $\delta$ Scuti stars are located in the lower part of the Instability Strip \citep{Rodriguez2000}, and therefore follow a period-luminosity (PL) relation that has been studied in detail over the years: since the beginning by \cite{Leavitt1912} who discovered the Classical Cepheids, up to \cite{McNamara2011}, \cite{Ziaali2019}, \cite{Poro-PL-2021} who updated the PL relation for $\delta$ Scutis. 
Although $\delta$ Sct stars can show variations on multiple periods, our data are not sensitive enough to detect them because of the short duration of the OW light curves.

\begin{figure}
\centering
\includegraphics[scale=0.55]{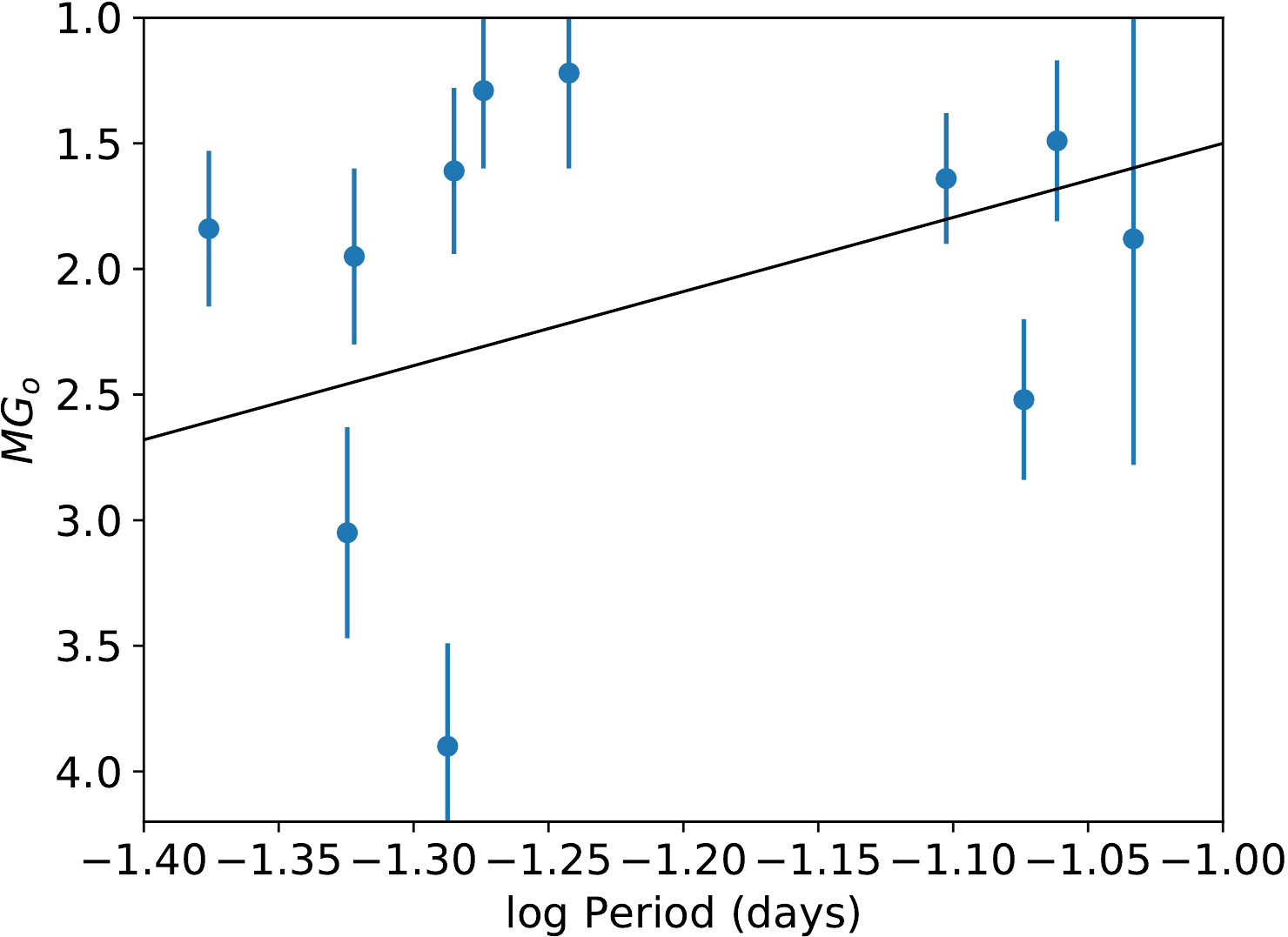}
\caption{The solid line shows the period luminosity relation of \citet{Ziaali2019} where we have converted $MV$ to $MG$ (see the text for details). The data points indicate the observed $MG_{o}$ for those stars which have evidence they are $\delta$ Sct stars which are OC members. Apart from one star (the lowermost point) there is reasonable agreement with the stars observed $MG_{o}$ and that predicted by the period luminosity relationship, thereby strengthening their classification as $\delta$ Sct stars.}
\label{periodlum}
\end{figure}

We now use the periods measured from the OW photometry to
determine the absolute magnitude of the $\delta$ Sct stars indicated
in Table \ref{clusterdeltaSct} using the PL relationship derived by \citet{Ziaali2019} based on Gaia DR2 data. In
addition we use the relationship between $G$ and $V$ using equation 1
of \citet{Montalto2021} to convert $M_{V}$ to $MG$. We estimate the
error on the $MG_{o}$ using the uncertainties on the distance estimate
from Gaia DR3 and add an additional 0.1 mag to take into account the
uncertainty of dereddening $MG$. We show the resulting predicted
values of $MG_{o}$ based on the period luminosity relationship in
Figure \ref{periodlum}.
The agreement between the predicted and measured values of $MG_{o}$ is reasonably good 
once the uncertainties are taken into account. The one
star which shows a significant departure from predictions is OW
J075556.7-275026.1 which we have classed as a probable cluster
member. We note that \citet{Poretti2005} find that $\delta$ Sct stars
which are pulsating in the first overtone rather than the fundamental
frequency have shorter periods by a factor of 0.775 which implies they
would be fainter by $\sim$ 0.3 mag: this would give a better fit to
the predicted absolute magnitude for around half the stars in Figure
\ref{periodlum}.

\begin{table*}
\centering
\caption{Those variable stars which are members or highly probable
  members of OCs. We note the cluster they are associated with, the OC
  age, the period of the modulation in the OW light curve, the variable star type, the likelihood of the star being an OC member and the
  dereddended Gaia absolute $G$ mag, $MG_{o}$, and $(B-R)_{o}$
  colour. Their light curves are shown in Figure~\ref{members-plot}.}
\begin{tabular}{lllrllrr}
\hline
  \multicolumn{1}{c}{Name} &
  \multicolumn{1}{c}{Cluster} &
  \multicolumn{1}{c}{log t} &
  \multicolumn{1}{c}{Period} &
  \multicolumn{1}{c}{Type} &
  \multicolumn{1}{c}{Members} &
  \multicolumn{1}{c}{$MG_{o}$} &
  \multicolumn{1}{c}{$(B-R)_{o}$}\\
  \multicolumn{1}{c}{} &
  \multicolumn{1}{c}{} &
  \multicolumn{1}{c}{(yr)} &
  \multicolumn{1}{c}{(min)} &
  \multicolumn{1}{c}{} &
  \multicolumn{1}{c}{} &
  \multicolumn{1}{c}{} &
  \multicolumn{1}{c}{} \\
  \hline
  OW J080237.1-294217.4 & ESO 430-09 & 8.5 & 70.6 & G? & prob & 7.71 & - \\
  OW J080659.2-300022.2 & NGC 2533 & 8.4 & 68.2 &  A8/F4 - $\delta$ Sct & Y & 3.05 & 0.60 \\
  OW J080713.1-295305.6 & NGC 2533 & 8.4 & 113.7 & $\delta$ Sct & Y & 1.64 & 0.59 \\
  OW J080706.9-295626.7 & NGC 2533 & 8.4 & 121.5  & A3 $\delta$ Sct & Y & 2.52 & 0.53 \\
  OW J080705.8-295140.8 & NGC 2533 & 8.4 & 133.4 & $\delta$ Sct & Y & 1.88 & 0.63 \\
  OW J075556.7-275026.1 & NGC 2483 & 7.6 & 74.3 &  $\delta$ Sct & prob & 3.90 & 0.88 \\
  OW J181531.5-220755.3 & NGC 6583 & 9.0 & 60.6 &  A7/A8 $\delta$ Sct & Y & 1.84 & 0.38 \\
  OW J181535.4-220934.6 & NGC 6583 & 9.0 & 68.6 &  $\delta$ Sct & Y & 1.95 & 0.50 \\
  OW J181542.9-220949.7 & NGC 6583 & 9.0 & 74.7 &  $\delta$ Sct & Y & 1.61 & 0.50 \\
  OW J181556.6-221021.4 & NGC 6583 & 9.0 & 76.6 &  $\delta$ Sct & Y & 1.29 & 0.63 \\
  OW J181555.0-220953.6 & NGC 6583 & 9.0 & 82.4 & $\delta$ Sct & Y & 1.22 & 0.53 \\
  OW J181545.5-220813.6 & NGC 6583 & 9.0 & 125.0 & $\delta$ Sct & Y & 1.49 & 0.66\\
\hline
\end{tabular}
\label{clusterdeltaSct}
\end{table*}

\section{Conclusions}

Using data obtained from the OW survey, we have presented the results
of a search for short period variable stars in the field of open clusters. The
data set (covering 134 deg$^2$ in the Galactic Plane and Bulge) was
cross-matched with the MWSC catalogue
\citep{Kharchenko-MWSCSurvey2013}, and catalogues based on Gaia
data (\citealt{Cantat-Gaudin-OCcat-GAIA-DR2-2018},
\citeyear{Cantat-Gaudin-OCcat-GAIA-DR2-2020}) to find stars which are
members of OCs. In a sample of 20 open clusters, we found 92 variable
stars of which 12 are OC members, which have a range of
ages. Based on spectroscopic observations and their location on the Gaia HRD, 12 of the variables are $\delta$ Sct pulsators. Of these, only 3 are physical members of open clusters. 

The results from Gaia are allowing the membership of OCs to be much
more reliably determined, and for new clusters to be identified
(e.g. \citealt{Ferreira2021}). This coupled with observations from space
missions such as {\sl Kepler} and {\sl TESS}
(e.g. \citealt{Nardiello2020}) and ground based surveys such as {\sl
  NGTS} (e.g. \citealt{Gillen2020}) or {\sl ZTF}
(e.g. \citealt{Coughlin2021}) will allow the photometric properties of
cluster members to be studied without the contamination of field
stars. In turn this will reveal how the properties of variable stars
vary as a function of age in a direct manner.
   
\section{ACKNOWLEDGEMENTS}

The data used in this paper are based on observations made with ESO
Telescopes at the La Silla Paranal Observatory under programme IDs 
088.D-4010(B), 090.D-0703(A), 090.D- 0703(B), 091.D-0716(A),
091.D-0716(B), 092.D-0853(B), 093.D- 0937(A), 093.D-0753(A),
094.D-0502(A), and 094.D-0502(B) as part of the Dutch GTO time on
OmegaCAM. This paper uses observations made at the South African
Astronomical Observatory (SAAO).  This research has made use of the
{\tt SIMBAD} database, operated at CDS, Strasbourg, France.  This
research has been facilitated by the NWO-NRF bilateral agreement for
astronomical collaboration between The Netherlands and the Republic of
South Africa. RT acknowledges funding from the project ``Big Data for Small Bodies'' (BD4SB),
this project being funded by a grant from the Romanian Ministry of Research and Innovation, CCCDI - UEFISCDI, project number PN-III-P1-1.1-TE-2019-1504, within PNCDI III.  
Armagh Observatory and Planetarium is core funded by the Northern Ireland Government. PW
and PJG acknowledge the Erasmus Mundus Programme SAPIENT, the National
Research Foundation of South Africa (NRF), the Nederlandse Organisatie
voor Wetenschappelijk Onderzoek (the Dutch Organization for Science
Research), Radboud University and the University of Cape Town.  RT
aknowledges the UK Royal National institute for Blind People -
Supporting people with sight loss for the offered Paid Work Placement
between August - October 2016, as part of the Eye Work Too Programme.
We thank the anonymous referee for a helpful report.

\section*{Data Availability}
The ESO data is available from the ESO archive. Light curves and
optical spectra which are shown in the paper can be requested from the
corresponding author.

\bibliographystyle{mnras}
\bibliography{ruxandra}  

\onecolumn
\appendix

\begin{landscape}
\section{The catalogue of variable stars}
 \label{Cat}
 
\scriptsize
 \centering
\begin{longtable}{rrrrrllllrrrrrrrrlll}
\caption[]{{\footnotesize The catalogue of variable stars found in the
    angular areas of a set of 20 open clusters overlapping OmegaWhite
    fields. We show the stars distributed in each cluster and sort
    them according to their period measured by the Lomb Scargle
    periodogram ($P_{LS}$). The star survey ID (that indicates the sky coordinates in hexadecimals, J2000) is
    given first, followed by the period and false alarm
    probability. For convenience and brevity, FAP is a notation used
    for log$_{10}$FAP here (See Paper 2 for details). The OmegaWhite
    photometry and the VPHAS+--DR2 (\citealt{DrewVPHAS+2014},
    \citeyear{Drew-VPHAS-DR2cat-2016}) $ugriH\alpha$ colour indices are given
    afterwards. The kinematic membership probability obtained
    from the MWSC catalogue \citep{Kharchenko-MWSCSurvey2013}, the
    membership probability obtained from Gaia DR2 data by
    \cite{Cantat-Gaudin-OCcat-GAIA-DR2-2018}, the UCAC4 memberships given by \cite{Sampedro2017}, the distance obtained from Gaia data by \cite{BailerJones2021}, the Gaia photometry \citep{Gaia2021}, 
    the membership based on the distance, the total membership and an estimated type of
    variable star are given in the last columns.}} 
\label{realvars} \\
\hline
  \multicolumn{1}{c}{ID} &
  \multicolumn{1}{r}{$P_{LS}$} &
  \multicolumn{1}{r}{FAP} &
  \multicolumn{1}{r}{OW$g$} &
  \multicolumn{1}{r}{{Amp}} &
  \multicolumn{1}{l}{$g-r$} &
  \multicolumn{1}{l}{$u-g$} &
\multicolumn{1}{l}{$r-i$} &
  \multicolumn{1}{l}{$r-H\alpha$} &
  \multicolumn{1}{r}{$P_{k}$} &
  \multicolumn{1}{r}{$P_{M}$} &
  \multicolumn{1}{r}{$P_{U4}$} &
  \multicolumn{1}{r}{d} &
  \multicolumn{1}{r}{$(BP-RP)_{o}$} &
  \multicolumn{1}{r}{$MG_{o}$} &
  \multicolumn{1}{l}{Mem} &
  \multicolumn{1}{l}{Overall} &  
  \multicolumn{1}{l}{Type}\\
     & (min) &     & (mag) & (mag) &(mag) & (mag) & (mag) & (mag) & ($\%$) & ($\%$) & & (pc) & & & Dist? & Mem? & \\
 \hline 
\endfirsthead   

\multicolumn{20}{c}{{\bfseries \tablename\ \thetable{} -- continued from previous page}} \\
\hline 
  \multicolumn{1}{c}{ID} &
  \multicolumn{1}{r}{$P_{LS}$} &
  \multicolumn{1}{r}{FAP} &
  \multicolumn{1}{r}{OW$g$} &
  \multicolumn{1}{r}{Amp} &
  \multicolumn{1}{r}{$g-r$} &
  \multicolumn{1}{r}{$u-g$} &
 \multicolumn{1}{r}{$r-i$} &
  \multicolumn{1}{r}{$r-H\alpha$} &
  \multicolumn{1}{r}{$P_{k}$} &
   \multicolumn{1}{r}{$P_{M}$} &
   \multicolumn{1}{r}{$P_{U4}$} &
   \multicolumn{1}{r}{d} &   
   \multicolumn{1}{r}{$(BP-RP)_{o}$} &
   \multicolumn{1}{r}{$MG_{o}$} &
   \multicolumn{1}{l}{Mem} &
     \multicolumn{1}{l}{Mem} &
   \multicolumn{1}{l}{Type} \\                                     
    & (min) &     & (mag) & (mag) &(mag) & (mag) & (mag) & (mag) & ($\%$) & ($\%$) & & (pc) & & & Dist? & Mem? & \\
\hline 
\endhead
\hline 
\multicolumn{20}{r}{{Continued on next page...}} \\ \hline
\endfoot
\\[-5pt] \hline \hline
\endlastfoot
\hline
\multicolumn{20}{l}{{ESO 430-09}}  \\
\hline      
OW J080237.1-294217.4  & 70.6 & -4.8 & 16.5 & 0.018   &   0.87  & 0.43   & 0.46   & 0.24    & 99.1 & & & 3090 & & 7.7 & Prob N & Prob & G? \\
OW J080234.2-294852.6  & 72.8 & -3.2 & 16.8 & 0.015   &   0.97  & 0.47   & 0.46   & 0.22    & 25.9 & & & 3130 & 0.59 & 2.8 & Prob N & Prob N &  G? \\
\hline
\multicolumn{20}{l}{{NGC 2533}}  \\
\hline                 
OW J080659.2-300022.2 & 68.2  & -3.1 & 16.5 & 0.014 &   0.82   & 0.39   &  0.42  &         & 0.0  & 70.0 &  & 2690 & 0.59 & 3.0 &  & Y & reddened A8\\
OW J080711.9-294621.7 & 86.1  & -4.7 & 17.5 & 0.063 &   0.74   & 0.46   &  0.44  & 0.23    & 9.2  & 50.0 &  & 4650 & 0.65 & 2.8 & N & N & A7 $\delta$ Sct \\
OW J080721.2-294516.6 & 113.0 & -5.0 & 16.5 & 0.064 &   0.75   & 0.35   &  0.38  & 0.24    & 66.7 & 50.0 &  & 4960 & 0.58 & 1.7 & N & N & A8 $\delta$ Sct  \\
OW J080713.1-295305.6 & 113.7 & -2.6 & 14.6 & 0.018 &   0.62   & 0.45   &  0.34  & 0.23    & 11.3 & 90.0 & Y & 2680 & 0.59 & 1.6 & & Y &  $\delta$ Sct? \\
OW J080706.9-295626.7 & 121.5 & -4.5 & 15.5 & 0.023 &   0.66   & 0.35   &  0.35  & 0.15    & 78.4 & 80.0 & Y & 2630 & 0.53 & 2.5 & & Y & A3 $\delta$ Sct \\
OW J080705.8-295140.8 & 133.4 & -4.2 & 15.7 & 0.043 &          &        &        &         & 72.9 & 60.0 & Y & 3150 & 0.63 & 1.9 & & Y & $\delta$ Sct? \\
\hline
  \multicolumn{20}{l}{{NGC 2483}}  \\
   \hline                              
OW J075556.7-275026.1 &  74.3 & -3.8 & 16.5 & 0.144 &   0.8     & 0.33   & 0.4  & 0.27    & 96.8 & & & 2190 & 0.88 & 3.9 & Prob N & Prob &  $\delta$ Sct?  \\
OW J075531.4-275953.0 &  83.7 & -2.7 & 16.2 & 0.011 &           & 0.59   & 0.43 & 0.19   & 73.9 & & & 2520 & 0.68 & 2.9 & N & N & $\delta$ Sct? \\
OW J075554.5-280000.6 &  87.6 & -5.3 & 14.9 & 0.025 &   0.66    & 0.48   & 0.34 & 0.24   & 64.7 & & & 2380 & 0.61 & 2.1 & N & N & $\delta$ Sct?  \\
OW J075542.2-275754.5 & 121.5 & -5.2 & 18.6 & 0.294 &   1.45    & 0.98   & 0.59 &        &      & & & 3480 & 1.14 & 4.2 & & & W UMa?    \\
OW J075609.2-275655.3 & 124.2 & -4.1 & 16.7 & 0.101 &   0.65    & 0.45   & 0.39 & 0.22   & 56.8 & & & 3870 & 0.70 & 2.9 & N & N & $\delta$ Sct? \\
OW J075516.4-275227.5 & 128.4 & -3.7 & 19.2 & 0.287 &   1.27    & 1.04   & 0.57 & 0.39   &      & & & 3000 & 1.17 & 5.6 & N & N & W UMa?        \\
OW J075518.8-275554.3 & 132.9 & -3.4 & 16.9 & 0.021 &   0.71    & 0.64   & 0.43 & 0.19   & 0.0  & & & 3950 & 0.58 & 2.9 & N & N & $\delta$ Sct? \\  
   \hline
   \multicolumn{20}{l}{{ESO 520-20}}  \\
   \hline                                  
OW J174334.1-244707.3  & 71.1 & -2.8 & 14.3 & 0.008 &    0.94    & 0.56   & 0.75 & 0.26   & 6.4 & & & 760 & 0.40 & 2.5 & N & N &  \\
OW J174336.0-243928.9  & 97.8 & -3.0 & 15.7 & 0.009 &    1.03    & 0.44   & 0.72 & 0.26   & 0.5 & & & 1140 & 0.59 & 3.0 & N & N &  \\
\hline
   \multicolumn{20}{l}{{Teutsch 14a}}  \\
   \hline                                                
OW J180303.1-221032.4 & 71.3 & -3.0 & 17.5 & 0.022    &         &        & 0.97 & 0.33     & 81.7 & 20.0 & & 1960 & 0.91 & 1.9 & & Prob N & $\delta$ Sct? \\ 
OW J180319.0-215901.8 & 79.2 & -3.6 & 17.0 & 0.023    &         &        & 1.19 & 0.44     & 54.3 & 10.0 & & 3320 & 1.08 & -0.9 & N & N & A/F $\delta$ Sct? \\ 
OW J180403.1-221024.6 & 96.7 & -4.9 & 17.7 & 0.062    & 1.72    & 1.34   & 0.98 & 0.35     & 0.0  & 10.0 & & 2430 & 0.97 & 1.4 & & N & $\delta$ Sct? \\ 
OW J180400.3-220932.2 & 111.8 & -4.3 & 19.7 & 0.183   & 1.77    &        & 1.0 & 0.35      &      & 20.0 & & 1910 & 1.25 & 4.2 & & N & $\delta$ Sct? \\ 
OW J180407.3-220601.2 & 112.9 & -4.2 & 19.5 & 0.114   & 2.0     & 1.7    & 1.08 &          & 74.7 & 10.0 & & 1860 & 1.15 & 3.1 & & N & $\delta$ Sct? \\ 
OW J180321.9-220717.3 & 126.7 & -4.6 & 17.8 & 0.029   &         &        &   &             & 50.2 & 40.0 & & 1530 & 1.20 & 1.9 & & N & $\delta$ Sct? \\ 
OW J180259.5-220937.8 & 132.6 & -5.0 & 16.9 & 0.042   &         &        & 1.11 & 0.39     & 94.9 & 40.0 & & 2860 & 1.24 & -0.1 & & Prob N & $\delta$ Sct? \\ \hline 
  \multicolumn{20}{l}{{ASCC-93}}  \\
\hline                                                  
OW J180753.6-220904.4 & 29.8 & -4.3 & 14.2 & 0.012    & 0.46    &        &   &             & 65.5 & & Y & 1170 & 0.28 & 2.4 & N & Prob N & A4 $\delta$ Sct \\
\hline
 \multicolumn{20}{l}{{NGC 6573}} \\
\hline                       
OW J181329.3-220502.4 & 58.3 & -3.8 & 15.1 & 0.010    & 0.78    & 0.33   & 0.55 &          & 63.2  & & & 1380 & 0.60 & 3.0 & N & N &  A5 $\delta$ Sct \\  
OW J181325.9-220559.3 & 79.4 & -2.6 & 19.5 & 0.106    & 1.37    & 0.86   & 0.82 & 0.36     &       & & & 3360 & 0.95 & 3.9 & & &      \\
OW J181323.6-220633.4 & 82.0 & -2.8 & 20.3 & 0.273    & 1.5     & 1.32   & 0.86 & 0.4      &       & & & 3920 & 1.04 & 4.1 & & &     \\
\hline
\multicolumn{20}{l}{{NGC 6583}} \\
\hline
OW J181531.5-220755.3  & 60.6  & -4.8 & 15.5 & 0.013 &          &        & 0.6  & 0.2       &  14.1 & 90.0 & & 2080 & 0.38 & 1.8 & & Y & A7/A8 $\delta$ Sct \\  
OW J181538.4-221317.6  & 64.2  & -2.6 & 18.4 & 0.045 &          & 0.78   & 1.04  & 0.34     &  49.8 & 30.0 & & 3690 & 1.20 & 3.0 & & N & $\delta$ Sct? \\
OW J181602.3-220537.2  & 64.9  & -3.0 & 16.4 & 0.012 &  0.94    & 0.57   & 0.63  & 0.18     &       & 20.0 & & 3020 & 0.50 & 1.7 & Prob N & N & $\delta$ Sct?     \\
OW J181535.4-220934.6  & 68.6  & -3.6 & 15.5 & 0.009 &  0.81    & 0.56   & 0.6  & 0.2       &  18.4 & 100.0 & & 2280 & 0.50 & 2.0 & & Y & $\delta$ Sct \\
OW J181603.8-220421.9  & 72.4  & -3.4 & 17.5 & 0.023 &  1.11    & 0.76   & 0.72  & 0.21     &       & 10.0 & & 2980 & 0.70 & 2.6 & & N & $\delta$ Sct?  \\
OW J181542.9-220949.7  & 74.7  & -2.6 & 15.0 & 0.007 &  0.8     & 0.59   & 0.6  & 0.19      &  0.0  & 80.0 & & 2190 & 0.50 & 1.6 & & Y & $\delta$ Sct?     \\
OW J181556.6-221021.4  & 76.6  & -4.3 & 15.1 & 0.012 &  0.94    & 0.7    &   &  0.11        &  0.0  & 100.0 & Y & 2350 & 0.63 & 1.3 & & Y  & $\delta$ Sct?     \\
OW J181555.0-220953.6  & 82.4  & -4.5 & 15.6 & 0.011 &  0.99    & 0.76   & 0.7  & 0.22      &  0.0  & 70.0 & Y & 2570 & 0.53 & 1.2 & & Y & $\delta$ Sct?     \\
OW J181545.5-220813.6  & 125.0 & -4.4 & 15.4 & 0.037 &  0.94    & 0.72   & 0.67  & 0.21     &  0.2  & 100.0 & N & 2320 & 0.66 & 1.5 & & Y & $\delta$ Sct?     \\
\hline
   \multicolumn{20}{l}{{Antalova 3}} \\
\hline                                                     
OW J173014.7-321035.4 & 99.5 & -2.6 & 18.0 & 0.031  &  1.65    & 1.36   & 1.09 & 0.39    & 9.8 & & & 1920 & 1.62 & 3.9 & & N & $\delta$ Sct?     \\
\hline 
\multicolumn{20}{l}{{ESO 430-14}} \\ 
\hline
OW J080309.0-312707.3  &  63.4 & -3.5 & 16.5 & 0.030 &  0.54    & 0.4    & 0.36 & 0.17      & 83.8 & & & 4270 & 0.02 & 1.8 & N & N & $\delta$ Sct?     \\
OW J080343.1-312637.1  &  83.2 & -2.9 & 19.7 & 0.191 &  0.97    & 0.86   & 0.49 & 0.43      &      & & & 7580 & 0.82 & 2.8 & N & N & $\delta$ Sct?     \\
OW J080338.5-312937.3  &  91.0 & -3.4 & 19.6 & 0.341 &  0.68    & 0.16   & 0.45 & 0.17      &      & & & 8980 & -0.02 & 2.8 & N & N & $\delta$ Sct?     \\
OW J080340.2-312002.7  &  91.7 & -3.1 & 19.8 & 0.239 &  1.11    & 0.64   & 0.56 &           &      & & & 4960 & 0.63 & 4.2 & N & N & $\delta$ Sct?     \\
OW J080335.9-312941.0  &  97.8 & -5.0 & 18.1 & 0.166 &  0.57    & 0.17   & 0.4 & 0.23       &      & & & 6960 & -0.06 & 1.9 & N & N & $\delta$ Sct?     \\
OW J080349.1-313005.8  & 103.9 & -3.5 & 18.6 & 0.072 &  1.64    & 1.56   & 1.15 & 0.61      & 86.3 & & & 1000 & 2.02 & 6.4 & N & N &    \\
OW J080404.2-312130.6  & 110.7 & -3.5 & 21.3 & 0.689 &  1.01    &        & 0.59 & 0.43      &      & & & 9070 & 0.28 & 3.9 & N & N & W UMa?  \\
OW J080342.0-312157.6  & 112.9 & -2.9 & 18.1 & 0.039 &  0.89    & 0.21   & 0.51 & 0.28      & 0.0  & & & 3170 & 0.46 & 3.8 &   & N & $\delta$ Sct?     \\
OW J080345.4-312313.3  & 114.0 & -4.0 & 19.7 & 0.195 &  0.92    & 0.29   & 0.5 & 0.3        &      & & & 4980 & 0.49 & 4.2 & Prob N & Prob N & $\delta$ Sct?     \\
OW J080402.5-312116.7  & 124.9 & -5.1 & 15.9 & 0.051 &  0.64    & 0.23   & 0.4 & 0.18       & 75.6 & & & 3220 & 0.20 & 1.8 & & Prob N & $\delta$ Sct?     \\
OW J080341.5-312617.2  & 127.6 & -4.7 & 20.2 & 0.469 &  1.17    &        & 0.53 &           &      & & & 5010 & 0.65 & 4.6 & N & N &    \\
OW J080357.1-312743.7  & 129.0 & -5.2 & 15.2 & 0.021 &  1.02    & 0.74   & 0.54 & 0.42      & 79.0 & & & 1120 & 0.88 & 4.1 & N & N & W UMa?  \\
OW J080337.9-312248.7  & 130.4 & -3.8 & 17.2 & 0.031 &  0.72    & 0.14   & 0.44 & 0.25      & 47.3 & & & 3030 & 0.32 & 3.3 &  & N & $\delta$ Sct?     \\
OW J080400.3-312627.4  & 131.9 & -4.1 & 19.7 & 0.220 &  1.61    &        & 0.79 & 0.47      &      & & & 1820 & 1.30 & 6.4 &  & & W UMa?  \\
OW J080357.2-312112.3  & 131.9 & -4.8 & 16.5 & 0.017 &  0.62    & 0.24   & 0.39 & 0.18      &      & & & 3050 & 0.24 & 2.5 &  & & W UMa?  \\
\hline
\multicolumn{20}{l}{{FSR 1342}}  \\
\hline
OW J075730.0-312747.8 & 92.3  & -3.3 & 20.5 & 0.381  &  1.36    &        & 0.86 & 0.38       &       &  20.0 & & 4490 & 0.93 & 4.7 & & N & $\delta$ Sct?     \\
OW J075735.0-312802.1 & 128.1 & -4.5 & 16.6 & 0.033  &          & 0.87   & 0.67 & 0.35       &  86.6 &  40.0 & & 3530 & 0.87 & 1.6 & & Prob N & $\delta$ Sct?     \\
\hline  
   \multicolumn{20}{l}{{Riddle 5}}  \\
\hline
OW J073645.9-244022.0 & 114.2 & -2.7 & 21.1 & 0.446 &  1.89    &        & 0.83 & 0.4         &        & & & 2260 & 1.32 & 6.4 & & & W UMa?  \\
OW J073646.1-244016.9 & 119.2 & -4.5 & 20.0 & 0.297 &  1.61    & 0.18   & 1.64 & 0.97        &  0.0   & & & 270 & 2.52 & 10.5 & N & N & W UMa?  \\
\hline                               
\multicolumn{20}{l}{{ESO 430-18}} \\ 
\hline                                   
OW J080759.0-303833.3  &  42.8   &  -2.6  &  19.0  &   0.112   &         &        &   &          &        &  & & 4540 & 0.99 & 4.2 & N & N &    \\
OW J080603.7-305050.5  &  43.7   &  -2.7  &  17.5  &   0.016   & 0.64    & 0.32   & 0.44 & 0.23  &        &  & & 4480 & 0.53 & 3.1 & N & N & A8 $\delta$ Sct    \\
OW J080603.5-305142.5  &  44.6   &  -4.0  &  16.7  &   0.022   & 0.46    & 0.25   & 0.35 & 0.2   &  13.1  &  & & 4410 & 0.33 & 2.5 & N & N & A4 $\delta$ Sct     \\
OW J080554.1-305750.0  &  52.1   &  -4.2  &  17.1  &   0.024   & 0.56    & 0.43   & 0.36 & 0.17  &  0.0   &  & & 4810 & 0.48 & 2.6 & N & N & A6 $\delta$ Sct     \\
OW J080610.3-304739.1  &  54.6   &  -3.0  &  20.7  &   0.756   & 1.23    &        & 0.77 & 0.4   &        &  & & 4570 & 1.22 & 5.8 & N & N & W UMa?  \\
OW J080557.6-310024.3  &  56.2   &  -3.4  &  20.6  &   0.544   & 1.53    &        & 0.73 & 0.41  &        &  & & 2250 & 1.67 & 7.1 & N & N & W UMa?  \\ 
OW J080635.2-305741.5  &  57.2   &  -3.4  &  14.1  &   0.041   & 0.51    & 0.06   & 0.41 &       &  0.0   &  & & 1800 & 0.40 & 2.0 & N & N & A0 $\delta$ Sct     \\
OW J080658.9-305759.3  &  60.0   &  -2.6  &  17.4  &   0.017   &         &        &   &          &  42.2  &  & & 4860 & 0.48 & 2.7 & N & N &   \\ 
OW J080739.7-310319.7  &  60.9   &  -2.8  &  17.3  &   0.020   &         &        &   &          &        &  & & 6630 & 0.33 & 2.0 & N & N &   \\ 
OW J080715.2-304346.3  &  60.9   &  -4.9  &  14.5  &   0.014   &         &        &   &          &  65.8  &  & & 1940 & 0.42 & 2.4 & N & N &   \\
OW J080713.4-305003.7  &  61.2   &  -2.6  &  17.1  &   0.016   &         &        &   &          &  20.1  &  & & 3680 & 0.64 & 3.3 & N & N &   \\
OW J080703.8-303504.5  &  63.8   &  -3.3  &  17.6  &   0.024   &         &        &   &          &        &  & & 4950 & 0.47 & 2.9 & N & N &$\delta$ Sct?     \\
OW J080639.9-310006.5  &  63.8   &  -3.7  &  16.3  &   0.012   & 0.67    & 0.3    & 0.39 & 0.21  &  18.3  &  & & 2770 & 0.56 & 3.1 & N & N &   \\
OW J080607.0-304226.5  &  64.5   &  -3.1  &  20.7  &   2.719   & 1.28    &        &   &          &        &  & & 3050 & 1.42 & 6.5 & N & N &   \\
OW J080701.7-310233.0  &  65.9   &  -2.7  &  19.4  &   0.088   &         &        &   &          &        &  & & 2980 & 1.14 & 5.5 & N & N &   \\
OW J080555.5-305724.0  &  66.2   &  -3.0  &  19.7  &   0.114   & 0.84    & 0.31   & 0.52 & 0.36  &        &  & & 6940 & 0.66 & 3.7 & N & N &  G?    \\ 
OW J080633.7-310345.1  &  68.9   &  -3.5  &  20.4  &   0.306   & 1.55    &        & 0.79 & 0.44  &        &  & & 2360 & 1.40 & 6.6 & N & N &W UMa?  \\ 
OW J080735.2-303856.2  &  72.1   &  -2.9  &  15.0  &   0.013   &         &        &   &          &  34.8  &  & & 2840 & 0.71 & 1.9 & N & N &    \\
OW J080810.3-305419.3  &  81.8   &  -3.0  &  16.7  &   0.016   &         &        &   &          &        &  & & 4520 & 0.32 & 2.6 & N & N &    \\
OW J080611.6-305342.1  &  106.9  &  -2.9  &  16.5  &   0.020   & 0.71    & 0.32   & 0.46 & 0.29  &  3.0   &  & & 3210 & 0.68 & 3.0 & N & N &$\delta$ Sct? \\
OW J080635.7-305800.3  &  112.8  &  -4.6  &  20.2  &   0.410   & 1.64    &        &   &          &        &  & & 1580 & 1.56 & 7.2 & N & N &   \\
OW J080724.4-304202.6  &  123.1  &  -3.9  &  21.0  &   1.549   &         &        &   &          &        &  & & 1550 & 1.84 & 8.1 & N & N &   \\
OW J080643.6-310656.2  &  124.4  &  -4.1  &  20.1  &   0.360   & 1.02    & 0.33   & 0.64 & 0.28  &        &  & & 6750 & 0.85 & 4.5 & N & N &W UMa?  \\ 
OW J080608.8-305836.1  &  125.6  &  -3.9  &  20.7  &   0.506   &         &        & 0.72 &       &        &  &  & 6300 & 0.95 & 4.9 & N &N &    \\
OW J080717.0-305430.2  &  125.6  &  -4.8  &  19.2  &   0.250   &         &        &   &          &        &  &  & 3750 & 0.96 & 4.8 & N &N & W UMa?  \\
OW J080649.3-310443.9  &  125.6  &  -3.8  &  16.8  &   0.020   & 1.07    & 0.36   & 0.61 & 0.36  &  55.6  &  &  & 5090 & 0.42 & 2.5 & N & N &    \\
OW J080605.1-305631.4  &  126.4  &  -3.9  &  14.9  &   0.017   &         & 0.03   &   &          &  26.2  &  &  & 920 & 0.75 & 4.2 &    & N &    \\
OW J080815.2-305635.7  &  128.3  &  -3.7  &  14.8  &   0.012   &         &        &   &          &  85.7  &  &  & 3280 & 0.21 & 1.9 & N & N &    \\ 
OW J080701.4-305427.6  &  129.6  &  -4.1  &  15.8  &   0.017   &         &        &   &          &  44.2  &  &  Y & 3350 & 0.35 & 2.4 & N & N &    \\
OW J080717.4-305414.2  &  129.6  &  -3.6  &  15.9  &   0.012   &         &        &   &          &  2.0   &  & & 4190 & 0.43 & 2.0 & N & N &    \\
OW J080700.5-305426.7  &  131.0  &  -3.9  &  16.0  &   0.015   &         &        &   &          &  47.7  &  & & 2030 & 0.72 & 3.7 & N & N & $\delta$ Sct?     \\ 
OW J080606.3-305423.9  &  131.0  &  -3.9  &  15.7  &   0.017   & 0.67    & -0.01  & 0.43 & 0.3   &  62.1          &  &  Y & 3480 & 0.33 & 2.2 & N & N &    \\
  \hline
   \multicolumn{20}{l}{{Ruprecht 51}}  \\
  \hline
OW J080337.0-303735.8 & 40.9 & -3.3 & 17.4 & 0.017 &   0.68    & 0.24   & 0.42 & 0.27      &        & & & 5420 & 0.56 & 2.6 & N & N &    \\
OW J080331.2-303442.0 & 55.4 & -3.1 & 17.5 & 0.034 &   0.88    & 0.4    & 0.48 & 0.27      & 1.5    & & & 3320 & 0.82 & 3.8 & & N & $\delta$ Sct?     \\
OW J080344.5-303536.5 & 74.8 & -3.2 & 14.4 & 0.013 &           &        & 0.28 &           & 62.8   & & & 2290 & 0.42 & 2.0 & & N &    \\
\hline
\end{longtable}

\end{landscape}
\normalsize

\section{The light curves of variable stars}
\label{allvars_lcs}

\begin{figure*}
\centering
\includegraphics[scale=0.39]{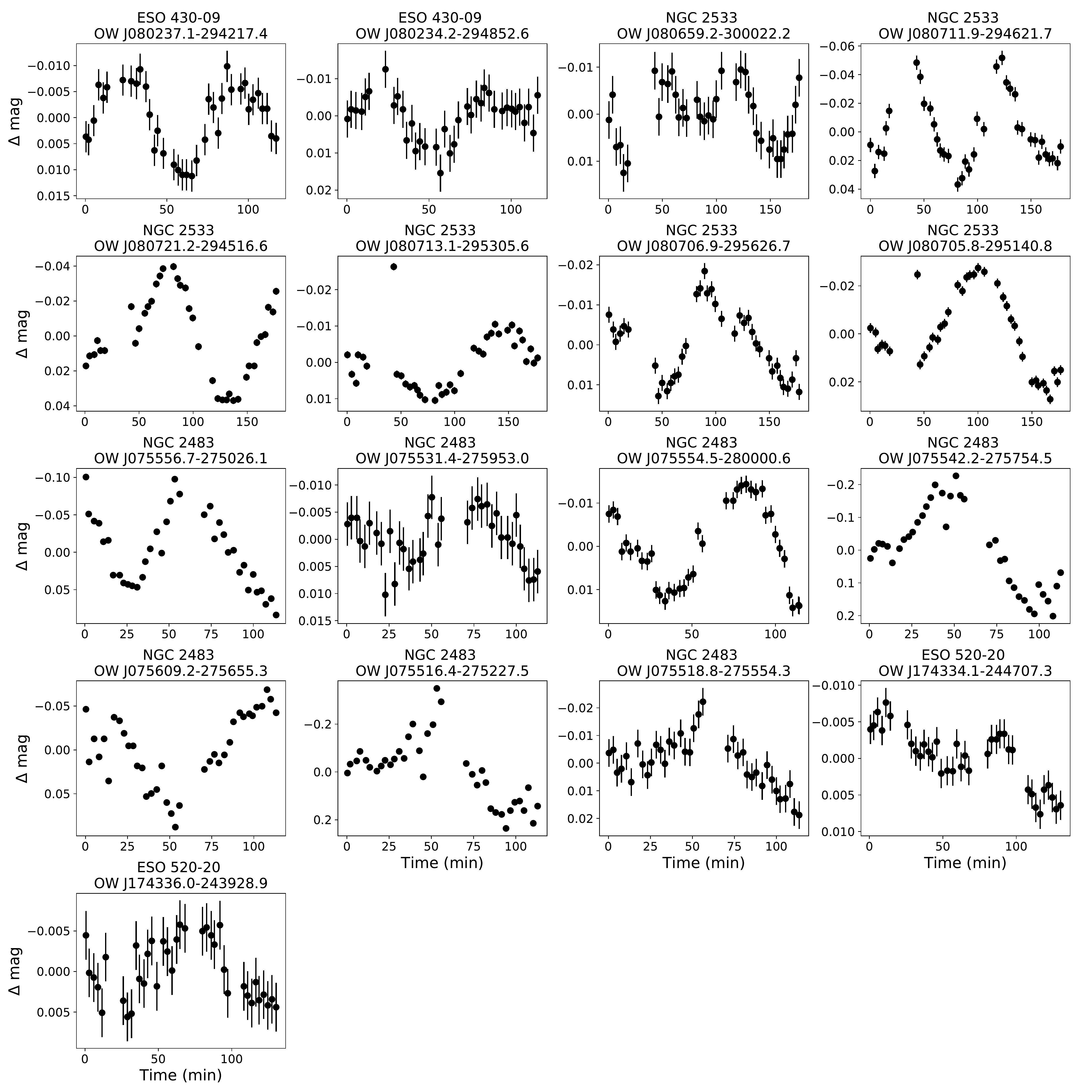}
\caption[Variable stars in clusters found in Sem 88, 91] {{\footnotesize The light curves of variable stars found in the angular areas of open clusters observed during the ESO Semesters 88 and 91. The stars are sorted according to their increasing periods, following the Table~\ref{realvars}. The cluster name and the survey ID are given in the title of each light curve}} 
\label{FIG:lcs1}
\end{figure*}

\begin{figure}
\centering
\includegraphics[scale=0.39]{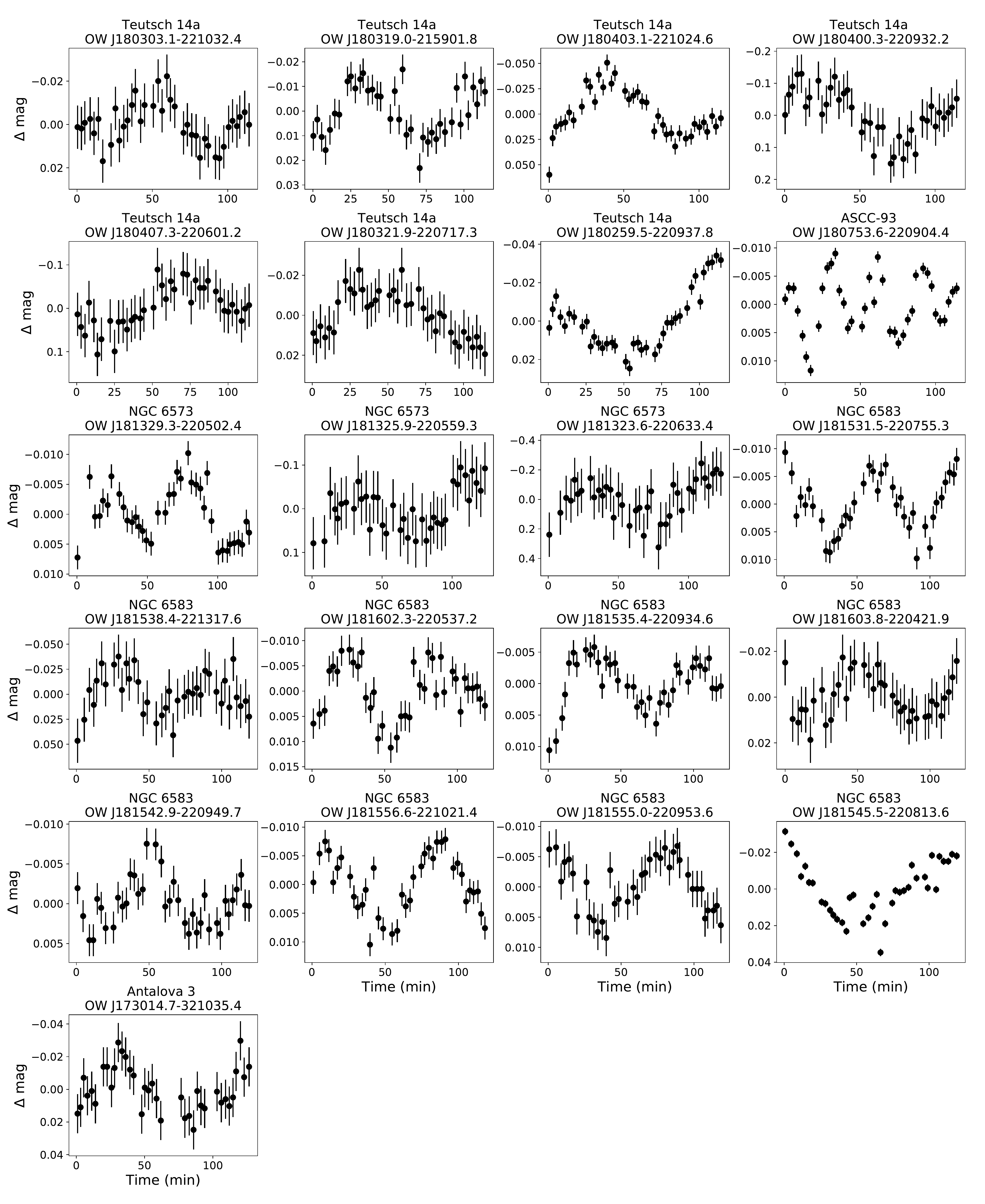}
\caption {As Fig.\,\ref{FIG:lcs1}  for open clusters observed during ESO semester 93.} 
\label{FIG:lcs2}
\end{figure}

\begin{figure}
\centering
\includegraphics[scale=0.39]{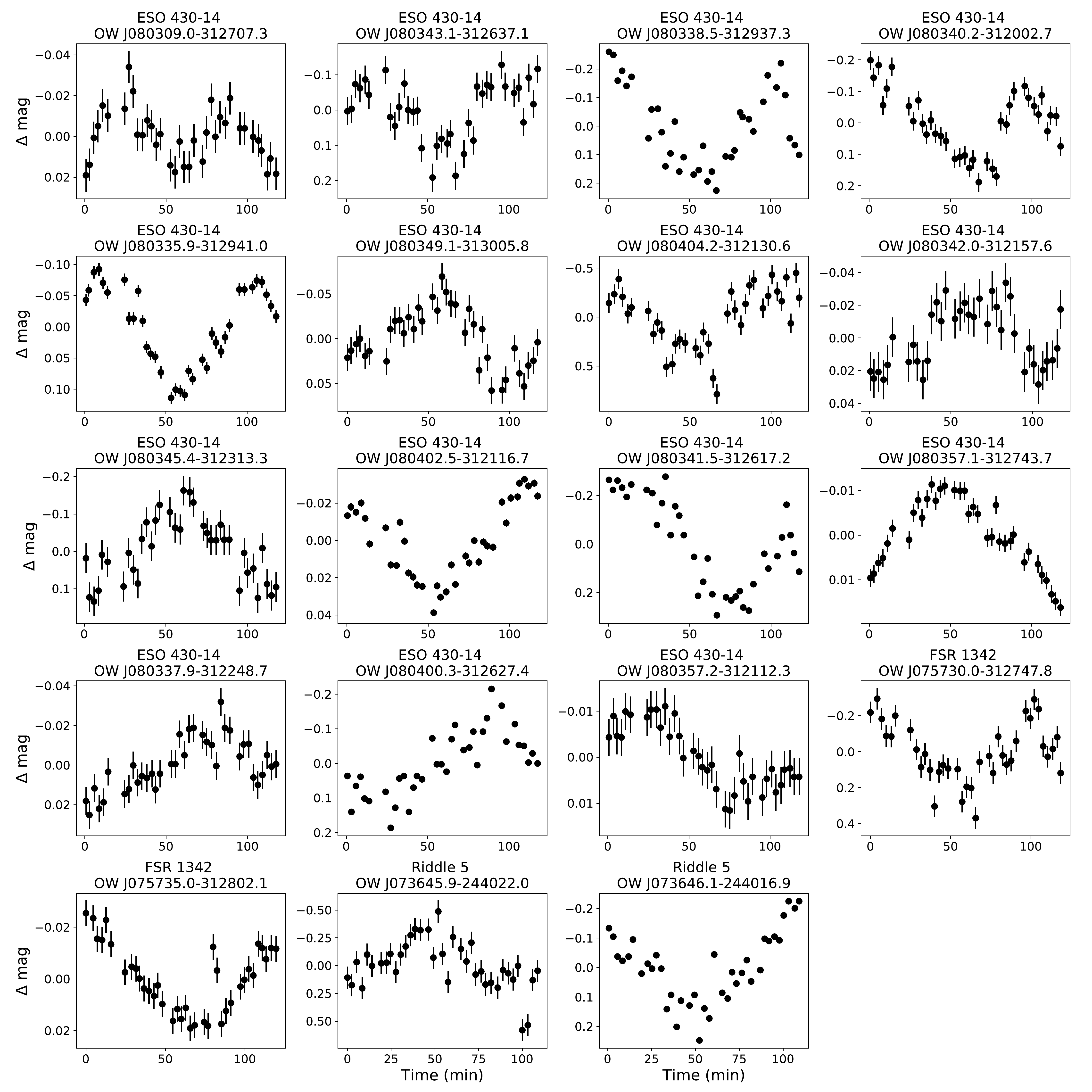}
\caption {As Fig.\,\ref{FIG:lcs1}  for open clusters observed during ESO semester 94.} 
\label{FIG:lcs3}
\end{figure}

\begin{figure}
\centering
\includegraphics[scale=0.39]{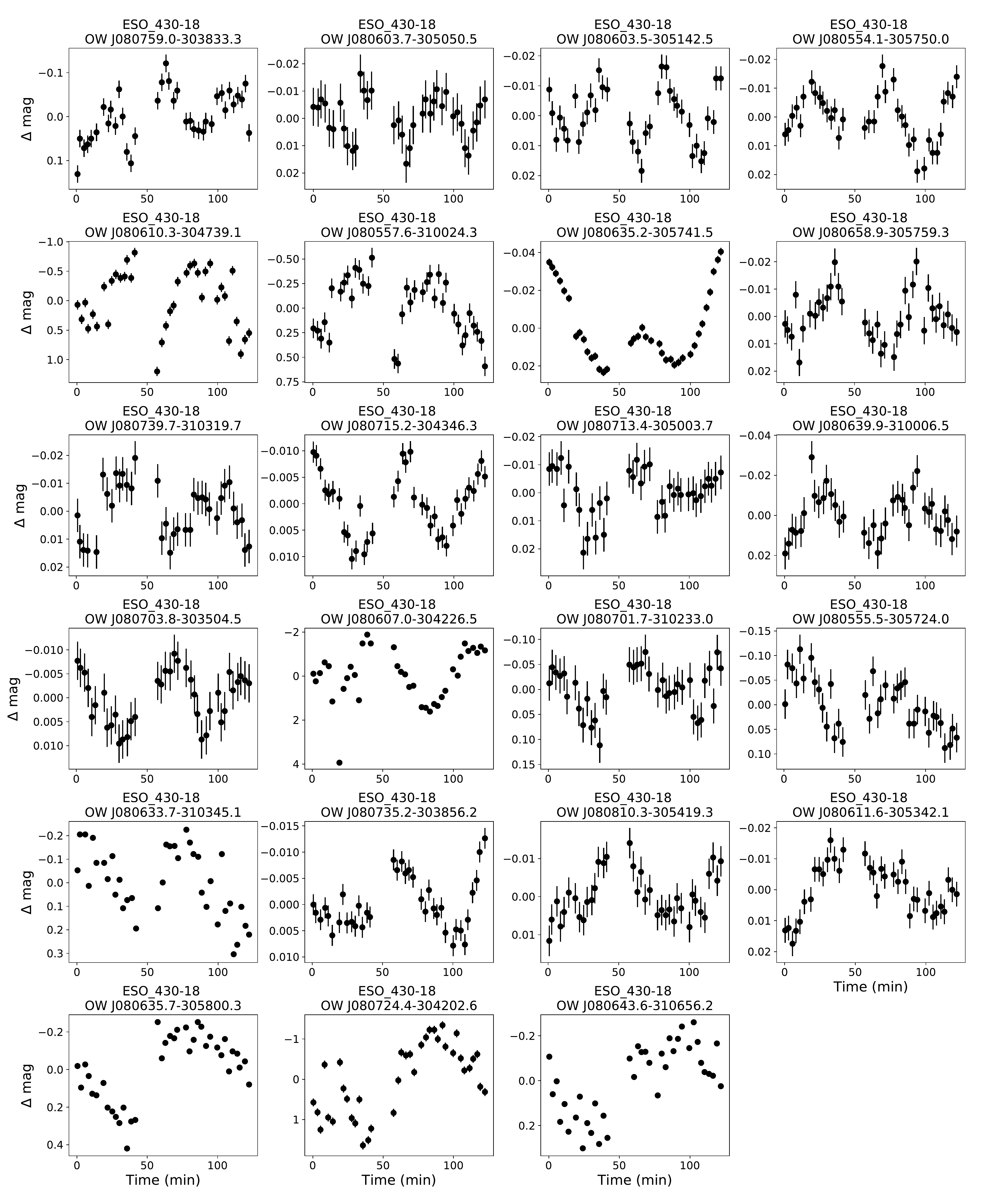}
\caption {As Fig.\,\ref{FIG:lcs1}  for open clusters observed during ESO semester 94 (cont).} 
\label{realvars-eso430-18-1}
\end{figure}

\begin{figure}
\centering
\includegraphics[scale=0.39]{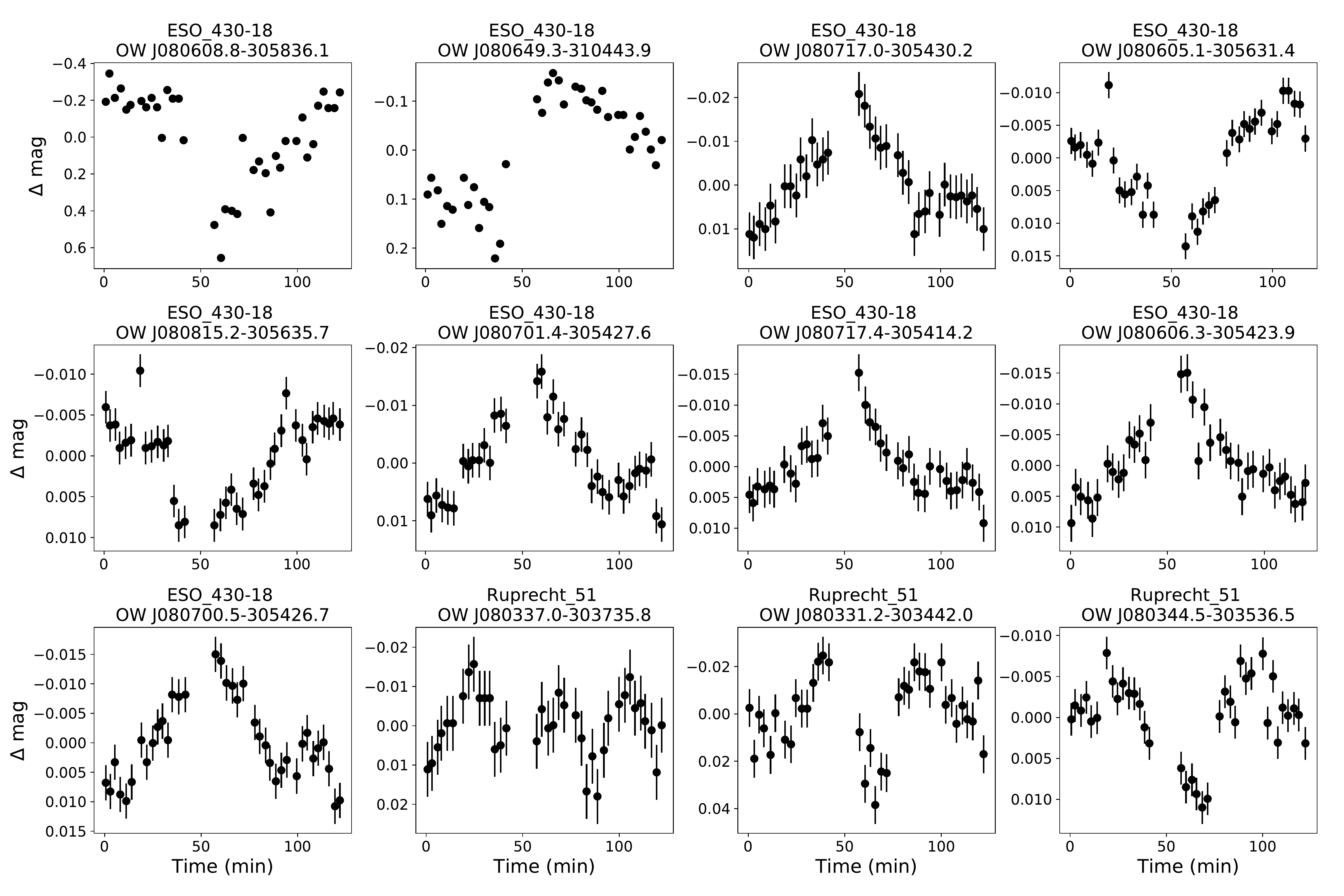}    
\caption {As Fig.\,\ref{FIG:lcs1}  for open clusters observed during ESO semester 94 (cont).} 
\label{realvars-eso430-18-2}
\end{figure}

\section{Identification of stellar types using the JHC library of spectra}
\label{ap-spectra-JHC}

\begin{figure}
\centering
\subfloat[Spectra of A vs F type stars - an example of comparison]{\label{FIG:sp-avsfstars}\includegraphics[scale=0.6]{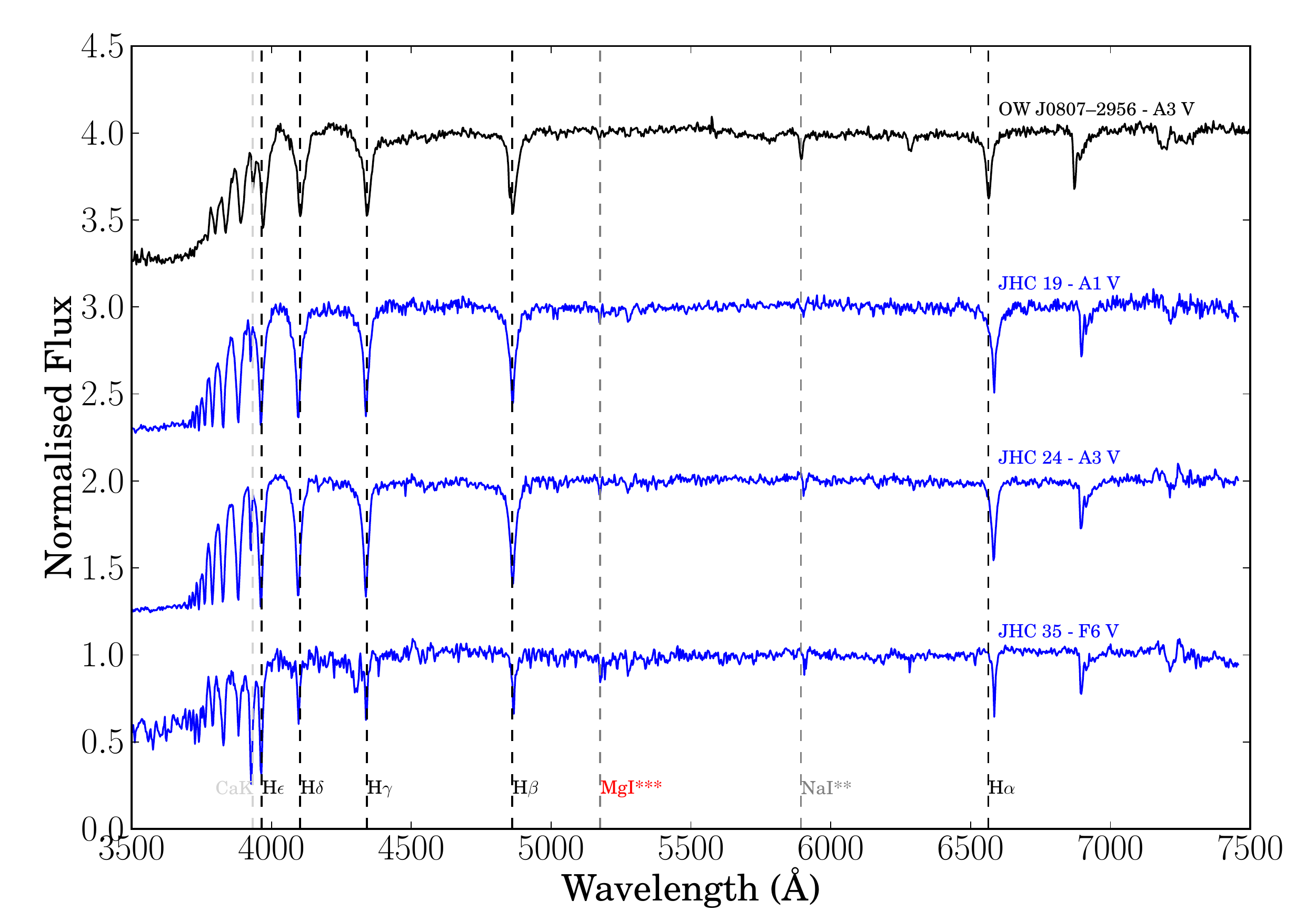}} 
\hspace{0.2cm}
\subfloat[Spectra of A type stars found in the area of open clusters.]{\label{FIG:sp-astars}\includegraphics[scale=0.6]{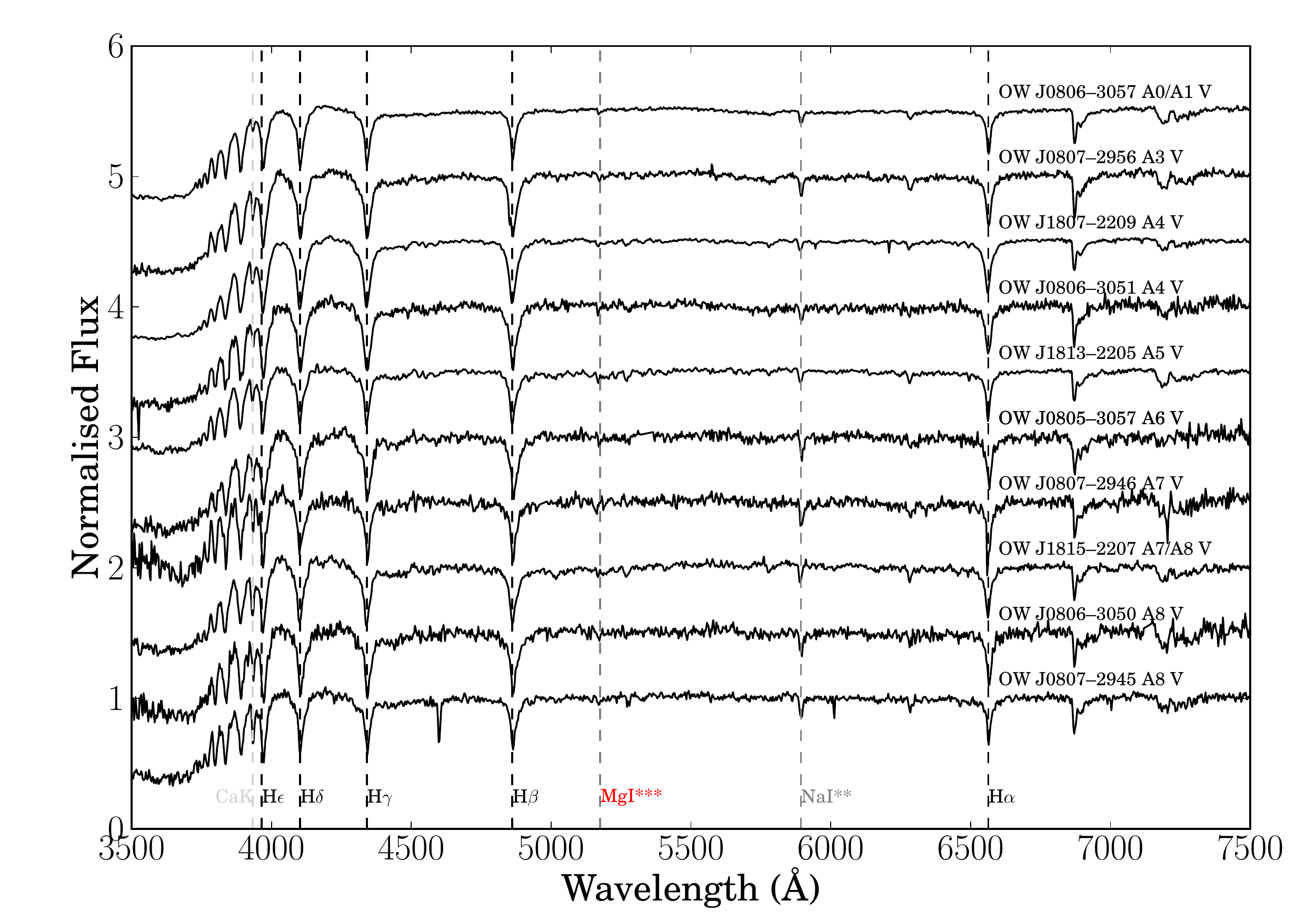}}  
\caption[]{{\footnotesize A set of spectra for variable stars observed in the angular area of 20 open clusters that overlap OmegaWhite fields. 
In the {\it top image}, we compare the spectrum of one OmegaWhite star (i.e. shown in black) with typical spectra of A and F stars (in blue) from the JHC atlas \citep{JHC-atlas1984} - as an example. Thus, we conclude that our source is an early A star. 
The {\it bottom figure} shows that ten of our stars are common A type stars. The membership probability for each star is given in Table~\ref{TAB:spectralID}.
See text for details.}}
\label{spectra1}
\end{figure}

\begin{figure}
\centering
{\label{FIG:sp-redstars}\includegraphics[scale = 0.6]{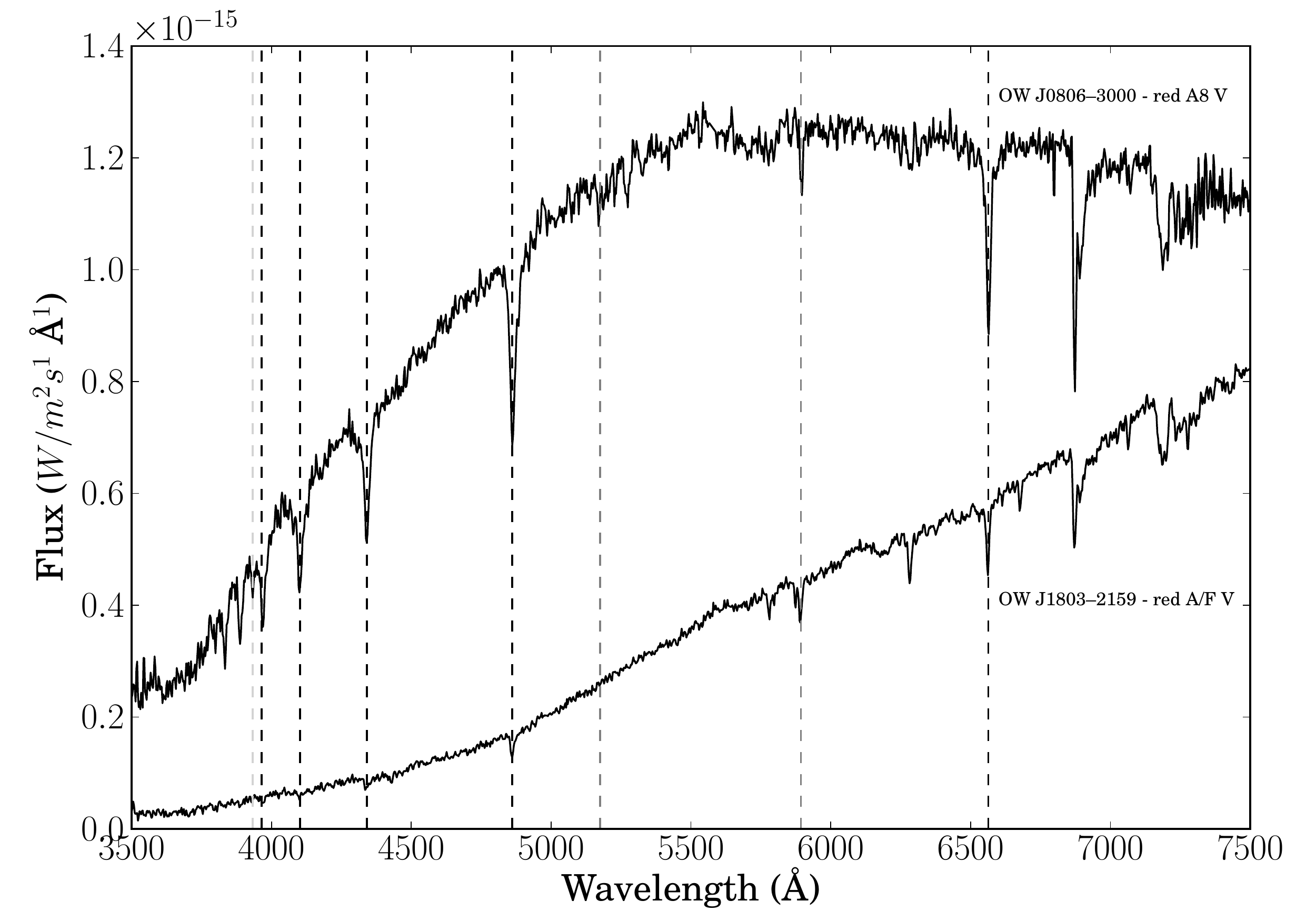}} 
\caption[]{{\footnotesize Examples of spectra of variable stars observed in the angular area of 20 open clusters that overlap OmegaWhite fields.
Two spectra of reddened A and A/F stars that look rather unusual
are displayed. See text for details.}}
\label{spectra2}
\end{figure}

\end{document}